\documentclass[review]{elsarticle}
\usepackage[mathscr]{euscript}
\usepackage[utf8]{inputenc} 
\usepackage[T1]{fontenc}    
\usepackage{url}            
\usepackage{float}
\floatstyle{plaintop}
\restylefloat{table}
\usepackage{multirow}
\usepackage{amsfonts}       
\usepackage{nicefrac}       
\usepackage{microtype}      
\usepackage{natbib}
\usepackage{amsmath}
\usepackage{color, soul}
\usepackage{breqn}
\usepackage{lineno}
\usepackage[margin=3cm]{geometry}
\usepackage{textcomp}
\usepackage{gensymb}
\usepackage{diagbox}
\usepackage{rotating}
\usepackage{enumerate}
\usepackage{caption} 
\usepackage{subcaption}
\usepackage{amssymb}
\usepackage{siunitx}
\usepackage{color, soul} 
\setulcolor{red}
\sethlcolor{yellow}
\renewcommand\hl[1]{#1} 

\usepackage{makecell}
\usepackage{longtable}
\usepackage{supertabular}
\usepackage{tabularx}
\usepackage{booktabs}
\usepackage{multirow}
\usepackage{multicol}
\usepackage{xltabular}
\usepackage{placeins}
\newcolumntype{Y}{>{\centering\arraybackslash}X}
\newcolumntype{Z}{>{\raggedright\arraybackslash}X}

\usepackage[pdftex,pagebackref=false]{hyperref} 
\hypersetup{
    plainpages=false,       
    unicode=false,          
    pdftoolbar=true,        
    pdfmenubar=true,        
    pdffitwindow=false,     
    pdfstartview={FitH},    
    pdfnewwindow=true,      
    colorlinks=true,        
    linkcolor=blue,         
    citecolor=green,        
    filecolor=magenta,      
    urlcolor=cyan           
}

\journal{ }

\begin{document}

\begin{frontmatter}

\title{The influence of beam focus during laser powder bed fusion of a high reflectivity aluminium alloy - AlSi10Mg}

\author[mymainaddress]{Sagar Patel}
\author[myTaddress]{Haoxiu Chen}
\author[mymainaddress]{Mihaela Vlasea \corref{mycorrespondingauthor}}
\cortext[mycorrespondingauthor]{}
\ead{mihaela.vlasea@uwaterloo.ca}
\author[myTaddress]{Yu Zou}

\address[mymainaddress]{University of Waterloo, Department of Mechanical and Mechatronics Engineering, Waterloo, ON N2L 3G1, Canada}
\address[myTaddress]{University of Toronto, Department of Materials Science and Engineering, Toronto, ON M5S3E4, Canada}

\begin{abstract}
The laser powder bed fusion (LPBF) of aluminium alloys is associated with numerous challenges when compared to other commonly used alloys (e.g., steels and titanium alloys) due to their higher reflectivity and thermal conductivity. This leads to a higher defect density in the final parts, commonly related to melt pool instabilities in the transition and keyhole melting modes. In this work, processing diagrams, temperature prediction models, X-ray computed tomography (XCT), and metallography are used for establishing criteria in process parameter optimization of high reflectivity aluminium alloys based on AlSi10Mg response in using 57 different power, velocity, and beam diameter combinations. For LPBF systems with focused beam diameters $<$100 \textmu m, divergent beams obtained by defocusing to a position above the LPBF build plate primarily lead to conduction mode melt pools, avoiding keyhole mode defects, and resulting in parts with densities >99.98\%, with effectively no porous defects in the subsurface regions. Additionally, an analytical model guided selection of laser power and velocity settings for a focused beam help in stabilizing melt pool and spatter dynamics in the transition melting mode thereby enabling a potential to obtain density values close to conduction mode densities ($\sim$99.98\%). A dimensionless keyhole number (Ke) was used in this work to identify distinct regions of conduction (Ke of 0-12), transition (Ke of 12-20), and keyhole (Ke $>$ 20) mode melting during LPBF of AlSi10Mg. Lastly, a melt pool aspect ratio (ratio of melt pool depth to width) of $\sim$0.4 is observed to be the threshold between conduction and transition/keyhole mode melt pools for AlSi10Mg, different from the conventionally assumed 0.5. This work demonstrates a dimensionless-process map method to obtain near fully dense parts that can be generalised for LPBF of high reflectivity alloys. 
\end{abstract}


\end{frontmatter}


\section{Introduction}
The potential of powder bed fusion (PBF) technologies to produce high-quality complex geometries and the ability to pursue assembly consolidation have been shown to increase technology adoption in the aviation \citep{BonninRoca2016Policy, Gisario2019Metal, Wagner2016Additive, Gorelik2017Additive, Najmon20192, Singamneni2019Additive, DebRoy2019Scientific} and automotive industries \cite{Delic2019Additive, Richardson2012Designer/maker, Hong2017Prospects, Delic2020effect, Leal2017Additive} in particular. Laser powder bed fusion (LPBF) is a PBF technology with a high industrial uptake, owing to the minimum feature size resolution and surface finish possible with the reduced beam spot sizes commonly used \cite{Gibson2014Additive}. The use of \hl{smaller} achievable beam spot sizes, combined with the advances in design for additive manufacturing (DfAM) competencies, have helped in realizing the true potential of LPBF for light weighting and design optimization of critical components. 

Aluminium alloys are commonly used in aerospace and automotive applications, when a combination of high strength performance and low weight is required. As such, AlSi10Mg has been adopted widely in the LPBF community for these applications \cite{Aboulkhair20193D, Brock2021Relative, Trevisan2017On}. The \hl{smaller} beam spot sizes used in LPBF, however, lead to the lowering of the vaporization threshold of aluminium \cite{Cunningham2019Keyhole}, which can be a disadvantage due to the ease of porosity formation in the keyhole melting mode of aluminium alloys \cite{Pastor1999porosity}. It has hence been previously assumed in literature that keyhole mode melting is the dominant melting mechanism in LPBF of aluminium alloys such as AlSi10Mg, while working with a beam spot radius of 10 \textmu m \cite{Aboulkhair2016On}. Similarly, for another experimentally intensive effort on processing AlSi10Mg with a beam spot radius of 35 \textmu m, keyhole mode porosity defects were observed in three-dimensional coupons at laser power settings ranging from 88 to 390 W, and scan speeds ranging from 250 mm/s to 2500 mm/s \cite{Brock2020laser}. Based on the literature, there are challenges in identifying stable process parameter windows for obtaining defect-free components for AlSi10Mg due to the rapid onset of keyhole melting mode driven by the interaction of a highly focused energy source acting on a high reflectivity and high thermal conductivity material system \cite{Aboulkhair2014Reducing, Thijs2013Fine-structured, patel2020melting}. The same can be said in terms of challenges in addressing porous defects in other aluminium alloys, such as A357 \cite{rao2016influence}, AlSi7Mg \cite{yang2018porosity}, Scalmalloy \cite{spierings2016microstructure}, Zr-modified Al6061 \cite{mehta2021additive}, and Zr-modified Al5083 \cite{zhou2019microstructure}. Simulations \cite{gan2021universal} and experiments \cite{Trapp2017In} \hl{to understand laser absorptivity during LPBF of aluminium have also pointed towards the significant differences in laser material interaction characteristics for aluminium alloys when compared to low reflectivity materials. High-speed and high-resolution X-ray imaging during LPBF of two aluminium alloys (AlSi10Mg and Al6061) has shown the formation of keyhole instability driven defects even at shallow melt pool depths due to the increased number of laser beam reflections in the vaporized region of Al melt pools} \cite{Hojjatzadeh2020Direct}. \hl{Additionally, melt pool vaporization has been shown to cause large spatter ejecta during LPBF} \cite{khairallah2020controlling} \hl{that is capable of blocking the laser beam leading to lack of fusion defects caused by a sudden drop in melt pool depth}. The significant differences in absorptivity characteristics for aluminium alloys in the conduction and transition melting modes when compared to titanium, ferrous, and nickel alloys \cite{Trapp2017In} also contribute towards the difficulties in obtaining defect-free parts as the rapidly increasing vapour cavity in the melt pool would make it difficult for a pore to escape during solidification. One path towards exploring such defect-free process windows is by exploring the effects of beam defocusing on achieving stable conduction-mode \cite{Pastor1999porosity, Metelkova2018On, Qi2017Selective}, as proposed in the present work. \hl{Stable conduction mode recipes also help in exploring faster beam velocities and higher hatch distances leading to productivity enhancement for LPBF} \cite{Metelkova2018On, tenbrock2020influence, sow2020influence}.

The motivation behind exploring defect-free stable conduction mode melting in AlSi10Mg is reinforced by the body of literature studying the effect of conduction, transition, and keyhole modes on microstructures, porosity and resulting mechanical properties of multiple alloy classes \cite{Qi2017Selective, tenbrock2020influence, Aggarwal2018Selective, Yang2016Role, Wang2019Microscale, patel2020towards}. Qi et al. \cite{Qi2017Selective} observe a lower crack density in the keyhole melting mode for Al7050 but a higher and more uniform nano-hardness across the melt pool in conduction melting mode. Higher vaporization of Zn and Mg was observed in the keyhole melting mode for Al7050. Aggarwal et al. \cite{Aggarwal2018Selective} observe a higher hardness, higher elongation, and finer cellular grains in stable keyhole mode coupons of 316L stainless steel, when compared to conduction mode coupons. Yang et al. \cite{Yang2016Role} report a wider processing window for Ti-6Al-4V in the conduction melting mode and similar tensile properties for conduction and keyhole melting modes, but report a higher elongation for keyhole melting mode coupons. Using micro-scale simulations, Wang and Zou \cite{Wang2019Microscale} report a more uniform thermal distribution during multi-track LPBF in the conduction mode when compared to keyhole melting mode, leading to a more uniform microstructure in Ti-6Al-4V. Patel et al. \cite{patel2020towards} reported reduced effects of adhered partially fused powder particles leading to low side-skin surface roughness for LPBF of Ti-6Al-4V using keyhole melting mode parameters when compared to conduction melting mode parameters. For a top-hat shaped laser beam profile, Tenbrock et al. \cite{tenbrock2020influence} reported $>$99.95\% density components in both the conduction and keyhole melting modes for 316L stainless steel with a gradual transition between the melting modes. A comparably uniform thermal distribution during multi-track printing leading to a uniform microstructure is hence a common characteristic observed in the stable conduction melting mode; with the potential for a finer microstructure, improved side-skin surface finish, and improved tensile properties for the stable keyhole melting mode. It is therefore important to explore methodologies for generating stable conduction mode and keyhole transition mode melting in aluminium alloys. 

In this work, a combination of processing diagrams, temperature prediction models, X-ray computed tomography (XCT), and metallography are used to study the effects of focused and divergent Gaussian laser beams during LPBF of 57 different process parameter combinations (21 using focused beams and 36 using divergent beams) during LPBF of AlSi10Mg. Dimensionless processing diagrams \cite{patel2020melting, gan2021universal} are used to identify distinct regions of conduction, transition, and keyhole melting modes during LPBF of AlSi10Mg. A temperature prediction model from previous work \cite{patel2020melting} is adapted for Al alloys with the goals of inferring the changes in laser absorptivity in the three melting modes. Additionally, an analytical model for predicting spatter dynamics is used for understanding and predicting high-density process parameter combinations using focused beams.

Deploying a process parameter strategy resulting in a drastic reduction in porous defects, specifically in the subsurface regions, \hl{would help improve part performance}, as porous defects are the most likely site for crack initiation \cite{Brandl2012Additive, Tang2017Oxides, Brandao2017Fatigue, Enrique2020Enhancing, Leuders2013On}; an example of a subsurface pore (diameter of $\sim$0.2 mm) leading to the fatigue crack in AlSi10Mg is shown by Plessis et al. \cite{Plessis2020Effects}. The approach presented in this work that uses experiments guided by normalized processing diagrams and temperature prediction models helps ease the process parameter optimization efforts for aluminium alloys with high reflectivity and high thermal conductivity, especially for the goals of reducing porosity. The results from this present work will help demonstrate that stable conduction mode melting can be achieved in high reflectivity and high thermal conductivity aluminium alloys, with a drastic reduction in core and sub-surface pore defects. The opportunity to obtain defect-free parts across melting modes during LPBF of aluminium alloys opens up new avenues in terms of tailoring microstructure for application-specific requirements due to significant differences in microstructure reported in literature across melting modes.   

\section{Materials and Methods}
A widely studied aluminium alloy, AlSi10Mg, was deployed using a modulated LPBF system (AM 400, Renishaw, UK) to demonstrate the theoretical concepts in this work. The powder used was plasma atomized pre-alloyed AlSi10Mg (AP\&C, Montreal, Canada) with a size distribution of 15-63 \textmu m and a D50 of 28 \textmu m. The experimental efforts in this work consists of 2 phases - melt pool analysis using weld lines and investigation of microstructure and porosity using cubes and cylinders respectively.

\subsection{Melt pool evaluation}
The first phase of this study is geared towards understanding the effects of the conduction, transition, and keyhole melting modes on the melt pool behaviour during LPBF of AlSi10Mg. As such, a total of 57 process parameter combinations were evaluated at varying laser powers (150 - 400 W), effective scanning velocities (250 - 1765 mm/s), beam spot radii (35 - 102 \textmu m), and powder layer thickness values (30 - 50 \textmu m) as summarized in Table~\ref{tab:AlSi10Mg_Process_Para}. 

For the modulated LPBF system used in the present study (AM 400, Renishaw, UK), the effective laser power, $P_{eff}$ is obtained from \cite{patel2020melting} and is given by Equation~\ref{P_eff}. Similarly, the effective beam velocity, $v$, is given by Equation~\ref{v_mod}. In Equation~\ref{P_eff}, $P$ is the actual laser power used in the modulated LPBF system, $t_e$ is the time when the laser is acting on the material (exposure time, \textmu s) and $t_d$ is the time when the laser is turned off and is re-positioning to the next exposure point (also referred to as the drill delay time, set as constant 10 \textmu s). In Equation~\ref{v_mod}, $p_d$ is the distance between two consecutive laser exposure points (point distance, \textmu m).

\begin{equation}
   P_{eff}=\frac{P \cdot t_e}{t_e+t_d}
   \label{P_eff}
\end{equation}

\begin{equation}
    v=\frac{p_d}{t_e+t_d}
   \label{v_mod}
\end{equation}

For the modulated LPBF system used in the present study (AM 400, Renishaw, UK), the beam spot radius at the focal point is given by $r_0 = 35$ \textmu m, and the wavelength of the laser beam used is $\lambda =$ 1070 nm. To obtain beam radius values > 35 \textmu m summarized in Table~\ref{tab:AlSi10Mg_Process_Para}, the laser beam was defocused to positions above the build plate, to obtain a divergent beam. The beam radius ($r_b$) values of the defocused beam are then obtained from the equation for a Gaussian distribution of a laser beam given by Equation~\ref{gaussian_laser} \cite{Bean2018Effect}.

\begin{equation}
    r_b=r_0 \sqrt{1+\left({\frac{z \lambda}{\pi r_0^2}}\right)^2}
   \label{gaussian_laser}
\end{equation}

\hl{For obtaining a beam radius of 54 \textmu m with the Renishaw AM 400, a defocusing distance of $z=4.2$ mm was used, $z=5.5$ mm was used to obtain a beam radius of 64 \textmu m, $z=6.6$ mm was used to obtain a beam radius of 73 \textmu m, $z=7.8$ mm was used to obtain a beam radius of 84 \textmu m, $z=8.7$ mm was used to obtain a beam radius of 92 \textmu m, and $z=9.9$ mm was used to obtain a beam radius of 102 \textmu m.}

\begin{xltabular}{\textwidth}{@{}YYYYYYYYY@{}}
\caption{LPBF processing parameters for evaluating the effect of melting modes on melt pool and porosity behaviour in AlSi10Mg. Laser power - $P$, point distance - $p_d$, exposure time - $t_e$, beam radius - $r_b$, layer thickness - $l_t$, hatch distance - $h_d$, effective laser power - $P_{eff}$, and effective velocity - $v$. \hl{The drill delay time, $t_d$, is constant for the Renishaw AM 400 at 10 \textmu s.}}
\label{tab:AlSi10Mg_Process_Para}\\
\toprule
    \textbf{\thead{Sample\\code}} & \textbf{\thead{$P$\\{}[W]}} & \textbf{\thead{$r_b$\\{}[\textmu m]}} & \textbf{\thead{$p_d$\\{}[\textmu m]}} & \textbf{\thead{$t_e$\\{}[\textmu s]}} & \textbf{\thead{$l_t$\\{}[\textmu m]}} & \textbf{\thead{$h_d$\\{}[\textmu m]}} & \textbf{\thead{$P_{eff}$\\{}[W]}} & \textbf{\thead{$v$\\{}[mm/s]}}\\
\midrule \midrule \endfirsthead
\multicolumn{9}{c}%
{\tablename\ \thetable{} -- Continued from previous page} \\
\toprule
    \textbf{\thead{Sample\\code}} & \textbf{\thead{$P$\\{}[W]}} & \textbf{\thead{$r_b$\\{}[\textmu m]}} & \textbf{\thead{$p_d$\\{}[\textmu m]}} & \textbf{\thead{$t_e$\\{}[\textmu m]}} & \textbf{\thead{$l_t$\\{}[\textmu m]}} & \textbf{\thead{$h_d$\\{}[\textmu m]}} & \textbf{\thead{$P_{eff}$\\{}[W]}} & \textbf{\thead{$v$\\{}[mm/s]}}\\
\midrule \midrule \endhead
       \hline \multicolumn{9}{|r|}{Continued on next page} \\
       \hline       \endfoot
       \hline       \endlastfoot
1 & 300 & 54 & 55 & 60 & 30 & 100 & 257 & 786 \\
2 & 350 & 54 & 55 & 80 & 30 & 100 & 311 & 611 \\
3 & 150 & 35 & 55 & 60 & 30 & 100 & 129 & 786 \\
4 & 180 & 35 & 55 & 70 & 30 & 100 & 158 & 688 \\
5 & 200 & 35 & 55 & 90 & 30 & 100 & 180 & 550 \\
6 & 240 & 35 & 55 & 100 & 30 & 100 & 218 & 500 \\
7 & 150 & 54 & 55 & 80 & 30 & 100 & 133 & 611 \\
8 & 200 & 54 & 55 & 100 & 30 & 100 & 182 & 500 \\
9 & 200 & 54 & 55 & 60 & 30 & 100 & 171 & 786 \\
10 & 250 & 54 & 55 & 80 & 30 & 100 & 222 & 611 \\
11 & 250 & 54 & 55 & 60 & 30 & 100 & 214 & 786 \\
12 & 300 & 54 & 55 & 100 & 30 & 100 & 273 & 500 \\
13 & 300 & 54 & 55 & 80 & 30 & 100 & 267 & 611 \\
14 & 350 & 54 & 55 & 60 & 30 & 100 & 300 & 786 \\
15 & 400 & 54 & 55 & 100 & 30 & 100 & 364 & 500 \\
16 & 400 & 54 & 55 & 80 & 30 & 100 & 356 & 611 \\
17 & 400 & 54 & 55 & 60 & 30 & 100 & 343 & 786 \\
18 & 200 & 64 & 55 & 60 & 30 & 100 & 171 & 786 \\
19 & 250 & 64 & 55 & 60 & 30 & 100 & 214 & 786 \\
20 & 300 & 64 & 55 & 60 & 30 & 100 & 257 & 786 \\
21 & 350 & 64 & 55 & 60 & 30 & 100 & 300 & 786 \\
22 & 400 & 64 & 55 & 60 & 30 & 100 & 343 & 786 \\
23 & 200 & 74 & 55 & 60 & 30 & 100 & 171 & 786 \\
24 & 250 & 74 & 55 & 60 & 30 & 100 & 214 & 786 \\
25 & 300 & 74 & 55 & 60 & 30 & 100 & 257 & 786 \\
26 & 350 & 74 & 55 & 60 & 30 & 100 & 300 & 786 \\
27 & 400 & 74 & 55 & 60 & 30 & 100 & 343 & 786 \\
28 & 200 & 84 & 55 & 60 & 30 & 100 & 171 & 786 \\
29 & 250 & 84 & 55 & 60 & 30 & 100 & 214 & 786 \\
30 & 300 & 84 & 55 & 60 & 30 & 100 & 257 & 786 \\
31 & 350 & 84 & 55 & 60 & 30 & 100 & 300 & 786 \\
32 & 400 & 84 & 55 & 60 & 30 & 100 & 343 & 786 \\
33 & 200 & 93 & 55 & 80 & 30 & 100 & 178 & 611 \\
34 & 250 & 93 & 55 & 80 & 30 & 100 & 222 & 611 \\
35 & 300 & 93 & 55 & 80 & 30 & 100 & 267 & 611 \\
36 & 350 & 93 & 55 & 80 & 30 & 100 & 311 & 611 \\
37 & 400 & 93 & 55 & 80 & 30 & 100 & 356 & 611 \\
38 & 300 & 102 & 55 & 100 & 30 & 100 & 273 & 500 \\
39 & 350 & 102 & 55 & 100 & 30 & 100 & 318 & 500 \\
40 & 400 & 102 & 55 & 100 & 30 & 100 & 364 & 500 \\
41 & 240 & 35 & 45 & 60 & 50 & 100 & 206 & 643 \\
42 & 240 & 35 & 60 & 60 & 50 & 100 & 206 & 857 \\
43 & 240 & 35 & 75 & 75 & 50 & 100 & 212 & 882 \\
44 & 200 & 35 & 60 & 170 & 50 & 100 & 189 & 333 \\
45 & 200 & 35 & 60 & 119 & 50 & 100 & 184 & 465 \\
46 & 200 & 35 & 60 & 60 & 50 & 100 & 171 & 857 \\
47 & 156 & 35 & 60 & 110 & 35 & 140 & 143 & 500 \\
48 & 193 & 35 & 60 & 70 & 35 & 140 & 169 & 750 \\
49 & 234 & 35 & 60 & 50 & 35 & 140 & 195 & 1000 \\
50 & 390 & 35 & 60 & 24 & 35 & 140 & 275 & 1765 \\
51 & 212 & 35 & 60 & 110 & 35 & 175 & 194 & 500 \\
52 & 270 & 35 & 60 & 70 & 35 & 175 & 236 & 750 \\
53 & 334 & 35 & 60 & 50 & 35 & 175 & 278 & 1000 \\
54 & 177 & 35 & 60 & 110 & 35 & 142 & 162 & 500 \\
55 & 193 & 35 & 60 & 70 & 35 & 140 & 169 & 750 \\
56 & 255 & 35 & 60 & 50 & 35 & 145 & 213 & 1000 \\
57 & 287 & 35 & 60 & 38 & 35 & 132 & 227 & 1250 \\
        \bottomrule
\end{xltabular}

\begin{figure}[htbp]
    \centering
    \captionsetup{justification=centering}
    \includegraphics[width=10cm,keepaspectratio]{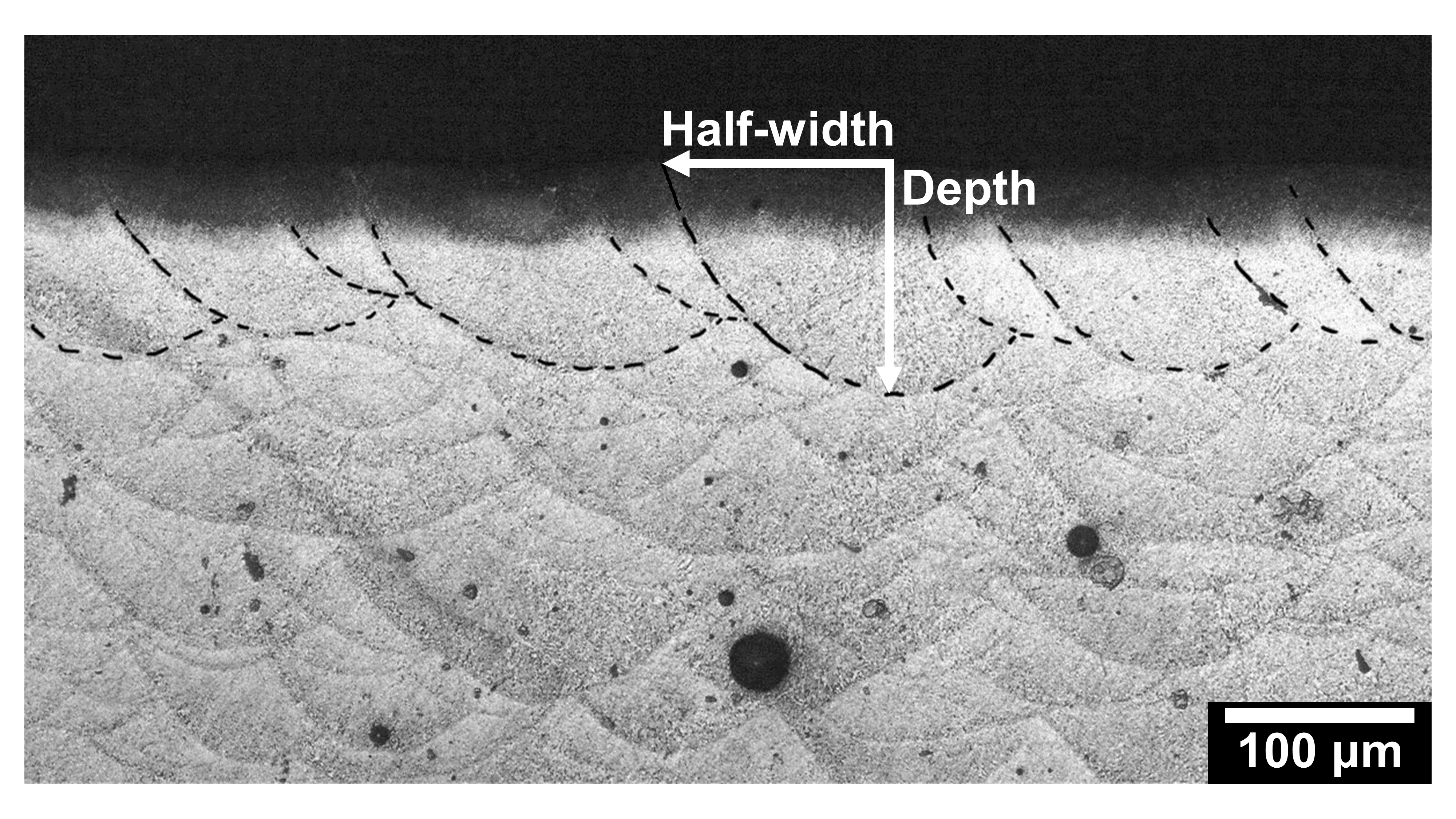}
    \caption{Measurement of the melt pool depth and half-width for one of the melt pools from sample 5.}
    \label{fig:measurement_AlSi10Mg_sample_E}
\end{figure}

From Table~\ref{tab:AlSi10Mg_Process_Para}, melt pool measurements for sample code labels 7-40 were obtained by printing 5 consecutive weld lines (length 8 mm, using the corresponding process parameters listed) on top of substrate artifacts. The substrate artifacts are cubes of side-length 10 mm which were printed directly on the reduced build volume of the AM 400. The 5 weld lines were separated by a hatch distance of 100 \textmu m. For these samples, meltpool measurements were obtained from the five weld lines, by cross-sectioning the samples perpendicular to the weld lines. The substrate artifacts were manufactured with a laser power of 200 W, point distance of 60 \textmu m, exposure time of 90 \textmu m, and beam spot radius of 54 \textmu m. For the remaining labels, cubes of side-length 10 mm were manufactured using the process parameters given in Table~\ref{tab:AlSi10Mg_Process_Para}, \hl{as part a separate print}, and a hatch distance of 100 \textmu m to obtain the melt pool depth and half width measurements as shown in Figure~\ref{fig:measurement_AlSi10Mg_sample_E}; such measurements were taken for meltpools where the meltpool depth is fully visible. For these samples, meltpool measurements were obtained from the top most layer, by cross-sectioning the samples perpendicular to the hatch vector on the top layer. The AlSi10Mg samples were sectioned, polished, and etched with diluted phosphoric acid (9 g phosphoric acid and 100 ml $\mathrm{H}_2\mathrm{O}$) for visualizing the desired meltpools. Micrographs were taken at the top edge of the cubes (VK-X250, Keyence, Japan).

\hl{The melt pool datasets are then analyzed using a normalized processing diagram, based on prior work} \cite{patel2020melting}, \hl{to evaluate the effect of divergent and focused beams on the melting modes during LPBF of AlSi10Mg using the 57 process parameter combinations from} Table~\ref{tab:AlSi10Mg_Process_Para}. \hl{$E^*$ and $v^*$ in the normalized processing diagrams are given by} Equations~\ref{E_star_AlSi10Mg} and \ref{v_star} respectively.

\begin{equation}
   E^*=\frac{A P_{eff}}{2 l_t \lambda (T_m - T_0)}
   \label{E_star_AlSi10Mg}
\end{equation}

\begin{equation}
   v^*=\frac{v r_b}{\alpha}
   \label{v_star}
\end{equation}

In Equation~\ref{E_star_AlSi10Mg}, \hl{$E^*$ is the dimensionless heat input, $A$ is laser absorptivity, $P_{eff}$ is the effective laser power [W], $l_t$ is the powder layer thickness [m], $\lambda$ is the thermal conductivity} [\si{W/(m K)}], \hl{$T_m$ is the melting temperature [K], and $T_0$ is the initial (or powder bed) temperature [K] taken as 293 K.} In Equation \ref{v_star}, \hl{$v^*$ is the dimensionless beam velocity, $v$ is the laser beam velocity [m/s], $r_b$ is the beam radius used [m], and $\alpha$ is the thermal diffusivity} [\si{m^{2}/s}]. The material properties used for Equations~\ref{E_star_AlSi10Mg} and \ref{v_star} are taken at the solidus temperature from \cite{Mukherjee2018Heat} and are given in Table~\ref{tab:AlSi10Mg_properties}. 

\begin{table}[htbp]
    \setlength{\extrarowheight}{2pt}
    \centering
    \captionsetup{justification=centering}
    \caption{Thermo-physical properties of AlSi10Mg taken at the solidus temperature \cite{Mukherjee2018Heat}.}
    \label{tab:AlSi10Mg_properties}
    \begin{tabular}{l c}
       \toprule
       \textbf{Properties} & \textbf{Material(AlSi10Mg)}\\
       \midrule \midrule
        Density, $\rho$ [\si{kg/m^{3}}]         & 2670 \\
         \midrule
        Thermal conductivity, $\lambda$ [\si{W/(m K)}] & 113  \\
         \midrule
        Specific heat capacity, $C_p$ [\si{J/(kg K)}]       & 565.29 \\
         \midrule
        Solidus temperature, $T_s$ [\si{K}]        & 831 \\
         \midrule
        Liquidus temperature, $T_m$ [\si{K}]       & 867 \\
         \midrule
        Vaporization temperature, $T_v$ [\si{K}]   & 2740 \\
        \midrule
        Total latent heat (fusion and vaporization), $H$ [\si{J/kg}] & 10943000 \\
       \bottomrule
    \end{tabular}
\end{table}

\hl{The threshold between the conduction and transition/keyhole melting modes during LPBF of AlSi10Mg in the processing diagram used in this work are given by the contour of the dimensionless peak temperature, $T_p^*$, based on previous work} \cite{patel2020melting}. \hl{The dimensionless peak temperature term, $T_p^*$, is given by} Equation~\ref{T_p_star_E}.

\begin{equation}
   T_p^*=\frac{T_p - T_0}{T_m - T_0}=\frac{3}{2 \sqrt{2} \pi e^{0.75}} \cdot \frac{E^*}{v^*} \cdot \frac{1}{(z^*+z_0^*)^2}
   \label{T_p_star_E}
\end{equation}

In Equation~\ref{T_p_star_E}, \hl{$T_p$ is the peak temperature under consideration, which is considered as the vaporization temperature (boiling point) of a given material, $T_m$ is the melting point of a given material, and $T_0$ is the powder bed temperature (which is room temperature for this work, 293 K). The term $z^*$ is dimensionless depth, obtained as the ratio of the dimension along the depth of a melt pool, $z$, and the beam spot radius, $r_b$, in} Equation~\ref{z_star}; this term was first defined by Ion et al. \cite{Ion1992Diagrams}. \hl{The term $z_0^*$ is the dimensionless distance of the apparent heat source above the surface of the melt pool, which is a function of $v^*$ as given by} Equation~\ref{z_star_0}. \hl{The detailed derivation for $T_p^*$ and associated terms is provided in previous work} \cite{patel2020melting}. 

\begin{equation}
   z^*=z/r_b
   \label{z_star}
\end{equation}

\begin{equation}
   z_0^*=\sqrt{\frac{3 \sqrt{\pi}}{2 \sqrt{2} e^{0.75}} \cdot \frac{1}{v^* tan^{-1}(\sqrt{8/v^*})}}
   \label{z_star_0}
\end{equation}

\hl{The threshold between the conduction and transition/keyhole melting modes for AlSi10Mg is given by the temperature contour for the boiling point of aluminium ($T_p^*=4.26$) at $z^*=0$. The term $z^*=0$ corresponds to surface vaporization and the contour in} Figure~\ref{fig:process_diagram_A_F_ALSi10Mg} \hl{is thereby labeled as the predicted surface vaporization threshold for LPBF of AlSi10Mg, or the threshold between the conduction and transition/keyhole melting modes.} 

\hl{Additionally, to affirm the findings by the normalized processing diagrams, as well as better identify the conduction, transition, and keyhole melting modes during LPBF of AlSi10Mg, a keyhole number ($Ke$) developed by Gan et al.} \cite{gan2021universal} \hl{is used in work. The keyhole number parameter, $Ke$, is given by} Equation~\ref{Ke_AlSi10Mg}.

\begin{equation}
   Ke=\frac{A P_{eff}}{\pi \rho C_p (T_m - T_0) \sqrt{\alpha v r_b^{3}}}
   \label{Ke_AlSi10Mg}
\end{equation}

\subsection{Microstructure and porosity evaluation} \label{microstructure-porosity}
The second phase of this study is geared towards understanding the effects of the conduction, transition, and keyhole melting modes on the microstructure and porosity during LPBF of AlSi10Mg. As such, cubes of side-length 10 mm (for microstructure evaluation) and cylinders of diameter 5 mm and height 9 mm (for porosity evaluation) were printed on the reduced build volume (RBV) of the AM 400 system. Six sets of processing parameters given by parameters sample code labels 1-6 (\hl{labels 1 and 2 from conduction mode, labels 3 and 4 from transition mode, and label 5 and 6 from keyhole mode}), from Table~\ref{tab:AlSi10Mg_Process_Para} were investigated. The hatching distance was kept constant at 100 \textmu m for all cubes and cylinders. The scan order was set such that the hatch volume (core) was scanned first, followed by the border; the border scans having the same set of processing parameters as the core. Border scans are commonly used in LPBF to improve the dimensional accuracy and surface roughness of LPBF coupons \cite{Chen2018Surface}; surface topography optimization was beyond the scope of this present work. The meander scan strategy was used with a 67° rotation between each layer, to reduce residual stresses, anisotropy, surface roughness and promote lower rates of defect propagation by the virtue of the scan vector direction not repeating for 180 layers \cite{Robinson2019effect, Dimter2008Method, Chen2020Grain}.

The AlSi10Mg cylinders were analyzed for porosity characteristics by a 3D X-ray computed tomography (XCT) scanner (ZEISS Xradia 520 Versa) using a 6 \textmu m voxel size. To visualize the defect distribution within each sample, the CT scanned files were analyzed using an image processing software (Dragonfly 3.0, Object Research Systems Inc., Montreal, QC). The AlSi10Mg cubes were sectioned, polished, and etched with diluted phosphoric acid (9 g phosphoric acid and 100 ml $\mathrm{H}_2\mathrm{O}$) for studying their microstructure. Micrographs were taken at various locations including the top edge and core of the cubes (VK-X250, Keyence, Japan). 

\section{Results and discussion}
In this work, laser beam defocusing was studied, primarily with the intent of understanding how to achieve low porosity parts via stable conduction mode and steady-state transition mode LPBF process parameters for AlSi10Mg. The literature discussed in detail in Section~\ref{melt_pool_melting_modes} indicates that during the LPBF of AlSi10Mg, conduction mode melting is not reported for beam spot radiuses below 50 \textmu m. Since the Renishaw AM 400 has a beam spot radius of 35 \textmu m at the focal plane, a transition or keyhole melting mode is hence expected, unless the beam is defocused. The defocusing of the beam was kept to positions above the build plate to create a divergent beam effect at the laser-material interaction plane, instead of a convergent beam effect obtained by defocusing to positions below the build plate.

\subsection{Melt pool evaluation based on melting modes} \label{melt_pool_melting_modes}
\begin{figure}[htbp]
    \centering
    \captionsetup{justification=centering}
    \includegraphics[width=13cm,keepaspectratio]{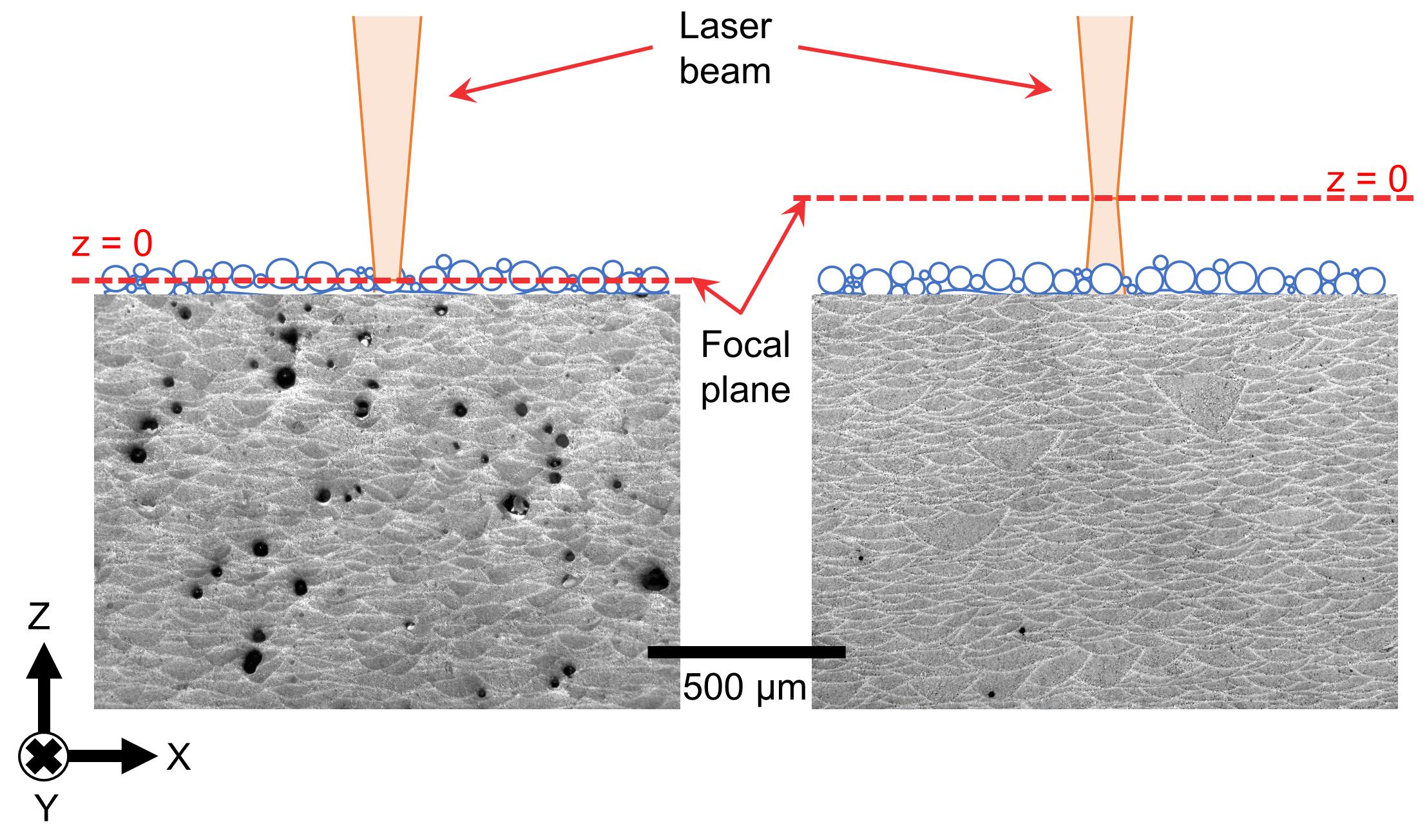}
    \caption{Typical microstructure of AlSi10Mg obtained with the focused beam (left, sample 5) and with the defocused beam (right, sample 2). The focused laser beam microstructure consists of numerous defects related to excessive vaporization in the keyhole mode, while a few hydrogen solubility related defects are observed by using a diverging defocused beam.}
    \label{fig:focused_divergent}
\end{figure}

The effect of the divergent beam defocusing strategy is illustrated in Figure~\ref{fig:focused_divergent}, wherein numerous large keyhole-type defects are generally observed for the samples built with the focused beam approach (for this study, a beam spot radius of 35 \textmu m), whereas a conduction mode type microstructure with a few minor defects are observed for the samples built with the divergent beam. Such minor defects captured in conduction mode melting of AlSi10Mg can be attributed to hydrogen-induced defects in aluminium alloys as reported by Weingarten et al. \cite{Weingarten2015Formation}. The melt pool outcomes, including the melt pool depths, width, and aspect ratios (melt pool depth/width) along with the values of dimensionless parameters used in this section are given in Table~\ref{tab:AlSi10Mg_Process_Outcomes}. 

\begin{figure}[htbp]
    \centering
    \captionsetup{justification=centering}
    \includegraphics[width=15cm,keepaspectratio]{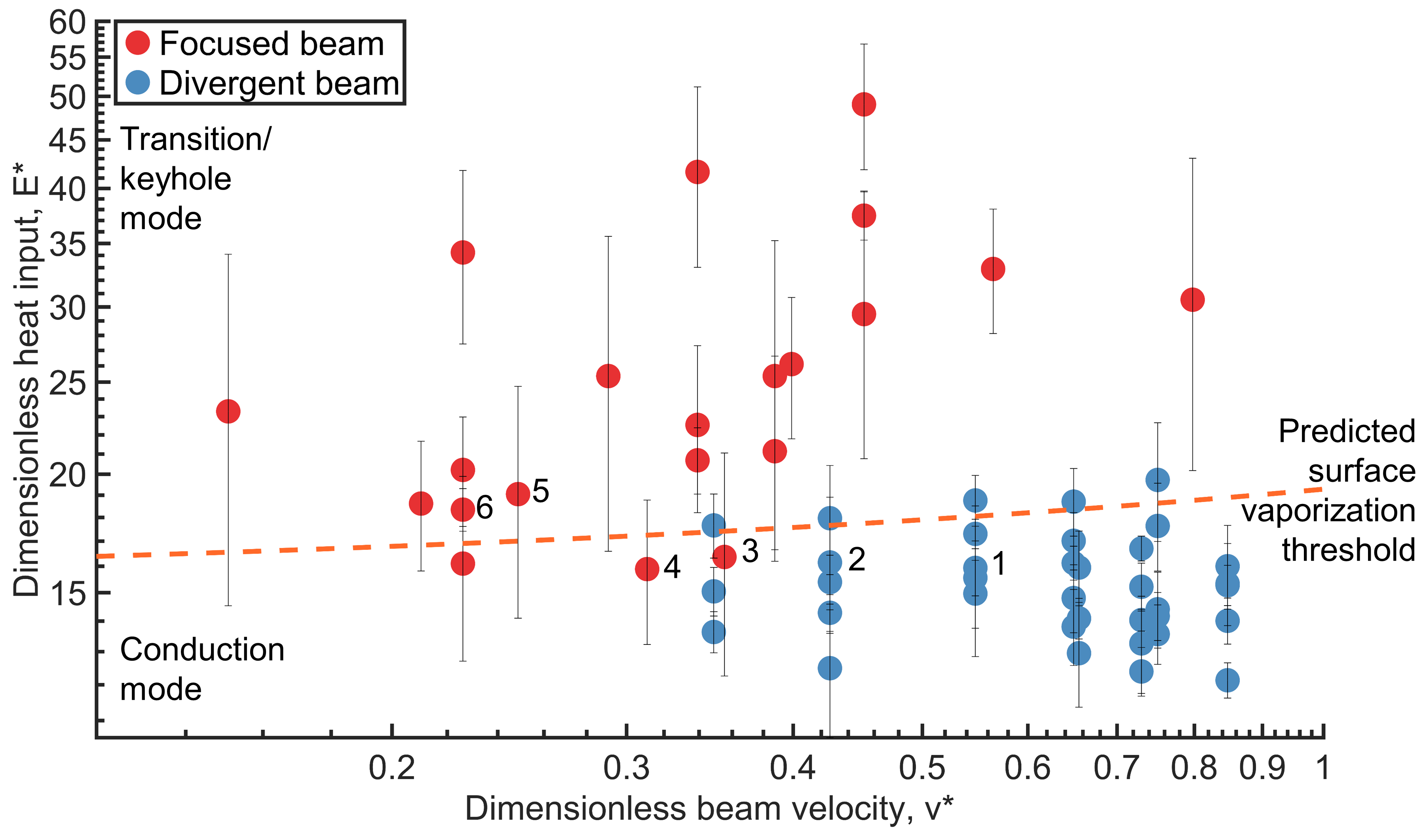}
    \caption{Normalized processing diagram for the 57 AlSi10Mg process parameters used in this study grouped by the use of focused and divergent beams. Samples codes 1-6 used for defect space evaluation in sections~\ref{defect_space_micrographs} and \ref{defect_space_XCT} are labelled in the figure.}
    \label{fig:process_diagram_A_F_ALSi10Mg}
\end{figure}

The normalized processing diagram illustrated in Figure~\ref{fig:process_diagram_A_F_ALSi10Mg} is used to evaluate the effect of divergent and focused beams on the melting modes during LPBF of AlSi10Mg using the 57 process parameter combinations from Table~\ref{tab:AlSi10Mg_Process_Para}. The absorptivity values used in the $E^*$ term (Equation~\ref{E_star_AlSi10Mg}) and given by the inferred absorptivity columns in Table~\ref{tab:AlSi10Mg_Process_Outcomes} were obtained inversely by comparing the predicted melt pool depths with experimental measurements (also given in Table~\ref{tab:AlSi10Mg_Process_Outcomes}). The average measured melt pool depth was used to inversely calculate the average inferred absorptivity, while the highest and lowest values within the confidence interval of the melt pool depth were used to calculate the variation in inferred absorptivity. \hl{The melt pool depth is substituted in the equation for $z^*$} (Equation~\ref{z_star}), \hl{and by setting a value of $T_p^*=1$, the absorptivity is calculated through the $E^*$ term in} Equation~\ref{T_p_star_E}. \hl{The calculation for absorptivity using this method now becomes a numerical problem, for which MATLAB's built-in 'vpasolve' function is deployed}. The values for the error bars for $E^*$ in Figure~\ref{fig:process_diagram_A_F_ALSi10Mg} are calculated by substituting the variation in inferred absorptivity into the $A$ term in Equation~\ref{E_star_AlSi10Mg}. 

\begin{xltabular}{\textwidth}{@{}cYYYYYYYYc@{}}
\caption{Melt pool outcomes (depth, width, and aspect ratios) and dimensionless parameters for the 57 process parameter combinations from Table~\ref{tab:AlSi10Mg_Process_Para}. Dimensionless heat input - $E^*$, dimensionless beam velocity - $v^*$, inferred laser absorptivity - $A$, keyhole number - $Ke$.}
\label{tab:AlSi10Mg_Process_Outcomes}\\
\toprule
    \textbf{\thead{Sample\\code}} & \textbf{\thead{$E^*$}} & \textbf{\thead{$v^*$}} & \textbf{\thead{$A$}} & \textbf{\thead{$Ke$}} & \multicolumn{2}{c}{\textbf{\thead{Melt pool\\depth (\textmu m)}}} & \multicolumn{2}{c}{\textbf{\thead{Melt pool\\width (\textmu m)}}} & \textbf{\thead{Aspect\\ratio}} \\
            \cmidrule(lr){6-9}
     & & & & &  \textbf{\thead{Avg.}} & \textbf{\thead{Std.\\dev.}} & \textbf{\thead{Avg.}} & \textbf{\thead{Std.\\dev.}} & \\    
\midrule \midrule \endfirsthead
\multicolumn{10}{c}%
{\tablename\ \thetable{} -- Continued from previous page} \\
\toprule
    \textbf{\thead{Sample\\code}} & \textbf{\thead{$E^*$}} & \textbf{\thead{$v^*$}} & \textbf{\thead{$A$}} & \textbf{\thead{$Ke$}} & \multicolumn{2}{c}{\textbf{\thead{Melt pool\\depth (\textmu m)}}} & \multicolumn{2}{c}{\textbf{\thead{Melt pool\\width (\textmu m)}}} & \textbf{\thead{Aspect\\ratio}} \\
            \cmidrule(lr){6-9}
     & & & & &  \textbf{\thead{Avg.}} & \textbf{\thead{Std.\\dev.}} & \textbf{\thead{Avg.}} & \textbf{\thead{Std.\\dev.}} & \\     
\midrule \midrule \endhead
       \hline \multicolumn{10}{|r|}{Continued on next page} \\
       \hline       \endfoot
       \hline       \endlastfoot
1 & 15.93 & 0.55 & 0.24 & 7.61 & 68.25 & 4.69 & 199.65 & 3.46 & 0.34 \\
2 & 16.14 & 0.43 & 0.2 & 8.75 & 79.35 & 3.26 & 213.1 & 23.83 & 0.37 \\
3 & 16.36 & 0.36 & 0.5 & 14.98 & 57.51 & 15.45 & 143.01 & 12.77 & 0.4 \\
4 & 15.89 & 0.31 & 0.39 & 15.55 & 59.98 & 10.58 & 136.05 & 17.05 & 0.44 \\
5 & 19.05 & 0.25 & 0.41 & 20.85 & 80.44 & 20.71 & 187.49 & 58.54 & 0.43 \\
6 & 18.35 & 0.23 & 0.33 & 21.06 & 81.67 & 3.95 & 170 & 35.14 & 0.48 \\
7 & 12.49 & 0.43 & 0.36 & 6.77 & 59.99 & 11.44 & 175.48 & 35.44 & 0.34 \\
8 & 13.64 & 0.35 & 0.29 & 8.17 & 74.04 & 4.08 & 205.47 & 25.63 & 0.36 \\
9 & 14.97 & 0.55 & 0.34 & 7.15 & 63.93 & 5.61 & 179.99 & 13.24 & 0.36 \\
10 & 14.29 & 0.43 & 0.25 & 7.74 & 69.83 & 3.37 & 201.64 & 7.92 & 0.35 \\
11 & 15.55 & 0.55 & 0.28 & 7.43 & 66.57 & 12.68 & 214.04 & 15.88 & 0.31 \\
12 & 15.05 & 0.35 & 0.21 & 9.01 & 82.3 & 5.06 & 223.11 & 13.74 & 0.37 \\
13 & 15.39 & 0.43 & 0.22 & 8.34 & 75.57 & 5.19 & 218.15 & 12.86 & 0.35 \\
14 & 17.3 & 0.55 & 0.22 & 8.27 & 74.19 & 2.62 & 204.36 & 15.77 & 0.36 \\
15 & 17.66 & 0.35 & 0.19 & 10.58 & 96.59 & 7.23 & 251.37 & 23.96 & 0.38 \\
16 & 17.97 & 0.43 & 0.2 & 9.74 & 88.21 & 11.22 & 256.91 & 38.46 & 0.34 \\
17 & 18.76 & 0.55 & 0.21 & 8.96 & 80.27 & 4.71 & 219.89 & 16.46 & 0.37 \\
18 & 13.82 & 0.65 & 0.31 & 5.12 & 63.1 & 6.58 & 213.2 & 17.61 & 0.3 \\
19 & 14.81 & 0.65 & 0.27 & 5.48 & 68.14 & 6.09 & 206.03 & 17.97 & 0.33 \\
20 & 16.12 & 0.65 & 0.24 & 5.97 & 74.56 & 3.18 & 215.94 & 6.6 & 0.35 \\
21 & 17.01 & 0.65 & 0.22 & 6.3 & 78.78 & 5.52 & 228.52 & 19.66 & 0.34 \\
22 & 18.71 & 0.65 & 0.21 & 6.93 & 86.49 & 6.81 & 246.73 & 5.4 & 0.35 \\
23 & 13.57 & 0.75 & 0.31 & 4.04 & 65.8 & 5.48 & 204.39 & 5.45 & 0.32 \\
24 & 14.17 & 0.75 & 0.26 & 4.22 & 69.16 & 4.53 & 213.35 & 7.86 & 0.32 \\
25 & 14.41 & 0.75 & 0.22 & 4.29 & 70.46 & 7.23 & 208.72 & 21.1 & 0.34 \\
26 & 17.64 & 0.75 & 0.23 & 5.25 & 87.14 & 9.24 & 254.41 & 22.05 & 0.34 \\
27 & 19.72 & 0.75 & 0.22 & 5.87 & 97.1 & 13.18 & 262.19 & 17.52 & 0.37 \\
28 & 12.13 & 0.85 & 0.28 & 3.01 & 60.47 & 3.26 & 206.97 & 22.57 & 0.29 \\
29 & 14.01 & 0.85 & 0.25 & 3.48 & 71.9 & 4.59 & 247.67 & 20.52 & 0.29 \\
30 & 15.34 & 0.85 & 0.23 & 3.81 & 79.5 & 8.55 & 267.62 & 12.59 & 0.3 \\
31 & 15.28 & 0.85 & 0.2 & 3.8 & 79.15 & 4.18 & 253.42 & 12.78 & 0.31 \\
32 & 16 & 0.85 & 0.18 & 3.97 & 83.15 & 8.9 & 249.12 & 21.4 & 0.33 \\
33 & 12.39 & 0.73 & 0.27 & 2.99 & 75.05 & 5.44 & 245.24 & 6.39 & 0.31 \\
34 & 13.27 & 0.73 & 0.23 & 3.21 & 81.43 & 11.24 & 238.8 & 17.81 & 0.34 \\
35 & 14.04 & 0.73 & 0.2 & 3.39 & 86.9 & 2.58 & 283.12 & 9.64 & 0.31 \\
36 & 15.22 & 0.73 & 0.19 & 3.68 & 95 & 5.94 & 297.45 & 24.59 & 0.32 \\
37 & 16.7 & 0.73 & 0.18 & 4.04 & 104.72 & 3.23 & 319.7 & 8.21 & 0.33 \\
38 & 12.95 & 0.66 & 0.18 & 3.01 & 92.41 & 13.85 & 322.46 & 22.89 & 0.29 \\
39 & 14.1 & 0.66 & 0.17 & 3.28 & 101.83 & 5.61 & 323.66 & 21.43 & 0.31 \\
40 & 15.95 & 0.66 & 0.17 & 3.71 & 116.31 & 10.97 & 354.77 & 7.05 & 0.33 \\
41 & 25.37 & 0.29 & 0.8 & 42.8 & 112.6 & 26.73 & 263.2 & 49.39 & 0.43 \\
42 & 25.37 & 0.39 & 0.8 & 37.07 & 112 & 20.58 & 245.2 & 34.72 & 0.46 \\
43 & 26.12 & 0.4 & 0.8 & 37.61 & 112.4 & 9.61 & 254.8 & 29.75 & 0.44 \\
44 & 23.29 & 0.15 & 0.8 & 54.56 & 124.6 & 44.22 & 244.8 & 72.84 & 0.51 \\
45 & 18.62 & 0.21 & 0.65 & 36.94 & 86 & 12.51 & 184 & 26.46 & 0.47 \\
46 & 21.14 & 0.39 & 0.8 & 30.89 & 82 & 14.09 & 195.6 & 19.36 & 0.42 \\
47 & 16.1 & 0.23 & 0.51 & 21.56 & 72 & 15.95 & 157.6 & 23.34 & 0.46 \\
48 & 22.54 & 0.34 & 0.61 & 24.64 & 79.4 & 13.94 & 178 & 29.9 & 0.45 \\
49 & 29.49 & 0.45 & 0.69 & 27.92 & 85.4 & 22.01 & 190.4 & 30.05 & 0.45 \\
50 & 30.53 & 0.8 & 0.5 & 21.76 & 65 & 19.61 & 141.2 & 22.39 & 0.46 \\
51 & 34.24 & 0.23 & 0.8 & 45.85 & 141.8 & 21.42 & 278.8 & 8.07 & 0.51 \\
52 & 41.63 & 0.34 & 0.8 & 45.51 & 133.4 & 19.98 & 238 & 23.79 & 0.56 \\
53 & 49.04 & 0.45 & 0.8 & 46.44 & 128.2 & 13.18 & 256.8 & 18.9 & 0.5 \\
54 & 20.21 & 0.23 & 0.57 & 27.06 & 89.2 & 10.62 & 189.6 & 20.76 & 0.47 \\
55 & 20.68 & 0.34 & 0.56 & 22.61 & 73.6 & 5.32 & 175.6 & 10.99 & 0.42 \\
56 & 37.44 & 0.45 & 0.8 & 35.45 & 110 & 4.3 & 205.2 & 20.08 & 0.54 \\
57 & 32.9 & 0.57 & 0.66 & 27.86 & 82.8 & 9.73 & 158.4 & 12.12 & 0.52 \\
        \bottomrule
\end{xltabular}       

An overwhelming majority of the sample codes using divergent beams are observed to lie beneath the predicted surface vaporization threshold (conduction mode), while the samples codes using a focused beam are generally observed to lie above the surface vaporization threshold (transition and keyhole modes), as illustrated in Figure~\ref{fig:process_diagram_A_F_ALSi10Mg}. The defocusing strategy is particularly relevant for LPBF systems with smaller beam spot radii (10-35 \textmu m) \cite{Aboulkhair2016On, Brock2020laser}, wherein transition and keyhole mode melt pools have been reported for all ranges of powers of 100 – 400 W for AlSi10Mg. For AlSi10Mg melt pool datasets reported with systems such as EOS M290 with a higher beam spot radius of 50 \textmu m at the focal point, most melt pools reported were observed to lie in the conduction melting mode, except for power settings above 275 W \cite{narra2017melt, scime2018Methods}. Divergent beams help in reducing the onset of keyhole and keyhole pores by reducing the effective beam power density as the melt pool formation progresses and have been used successfully in the laser welding of aluminum alloys \cite{Pastor1999porosity} and LPBF of 316L stainless steel \cite{Metelkova2018On}. \hl{When a divergent beam is used, a keyhole depression (if any) during LPBF moves away from the focal plane, leading to a decrease in power density as the melt pool formation progresses into the previously solidified metal. Thus, a divergent beam restricts a potential keyhole cavity from any further growth. On the other hand, a convergent beam exposes the keyhole cavity to an increasing powder density as melt pool formation progresses. Thus, even a shallow keyhole cavity is enough to grow into a deep keyhole and subsequently produce a large melt pool when a convergent beam is used, as shown by} Pastor et al. \cite{Pastor1999porosity} for laser welding, and Metelkova et al. for LPBF \cite{Metelkova2018On}. The cause of the observed absence of keyhole pores using divergent beams could be associated to a deviance of the beam profile from a Gaussian distribution to resemble more closely a top-hat distribution during the divergence of the beam as shown by Nie et al. \cite{Nie2020Effect}. Assuming this deviance of the beam profile, the threshold power required for surface vaporization (commonly assumed to be the threshold between the conduction and transition modes) would be higher for divergent beams when compared to focused beams, when all other variables are kept constant. This statement is also supported by the temperature prediction models proposed by Graf et al. \cite{Graf2015Analytical} for predicting the threshold of surface vaporization for Gaussian and top-hat beam profiles for materials with high thermal conductivity and low surface tension such as aluminium and copper alloys. Graf et al. \cite{Graf2015Analytical} show that in a Gaussian beam, when all other variables are held constant, a lower power is needed for initiating surface vaporization due to the higher peak intensity in Gaussian beam profiles, when compared to the top-hat distribution. For LPBF systems with beam spot radii <50 \textmu m, such as the Renishaw AM 400 used in this work, divergent laser beams are hence required to reduce the effect energy density and achieve conduction mode melt pools. 

The temperature prediction model used for Figure~\ref{fig:process_diagram_A_F_ALSi10Mg} has some limitations such as assumptions of a 2D heat source, temperature independent material properties, oversimplification of powder layer thickness effects, and ignorance of heat loss by refraction in the vapour plume \cite{patel2020melting}, which would contribute to the uncertainty margins in the identified surface vaporization threshold. In addition, the latent heat of fusion, thermo-capillary phenomena (Marangoni effect), recoil pressure, and varying laser power absorptivity due to the its angle of incidence (Brewster effect) are not incorporated into this modelling approach which could add to uncertainties \cite{Huang2016comprehensive, Zhang20193-Dimensional}. Additionally, 9 out of the 21 of the transition and keyhole mode melt pools had inferred absorptivity values >0.8, which are higher than experimentally measured limits of absorptivity during LPBF \cite{Trapp2017In, Matthews2018Direct, ye2019energy}. This is mainly due to the model limitations of not accounting for the effects of recoil pressure on the melt pool depth, which is particularly significant after the onset of surface vaporization in the transition and keyhole melting modes \cite{Khairallah2016Laser}. For these 9 sets of process parameters, an average absorptivity value of 0.8 was assumed. The use of standard deviation bars for $E^*$ as inferred inversely via the melt pool datasets are also a reflection of some of the limitations in experimentally validating the precise location of each experimental point in the process map. Despite the assumptions and limitations of the present model, such dimensionless process maps are of utility in experimental planning and in explaining experimental outcomes based on laser-material interaction phenomena.  

Recent attempts by Gan et al. \cite{gan2021universal} to develop universal scaling laws for keyhole porosity in LPBF make use of exponential absorptivity scaling laws for titanium, ferrous, and aluminium alloys. The absorptivty law used by Gan et al.  \cite{gan2021universal} was derived by Ye et al. \cite{ye2019energy} using in-situ micro-calorimetry measurements of absorptivity during LPBF of low reflectivity titanium, ferrous, and nickel alloys. While Gan et al. \cite{gan2021universal} use simulations to derive the same scaling law for high reflectivity aluminium alloys as well, the recommendations by Gan et al. \cite{gan2021universal} do not hold true for aluminium alloys when applied to the experimental melt pool datasets in this work, as well as previous work \cite{Brock2020laser}. The inferred absorptivity calculations from melt pool data, as proposed in this work follow the experimentally measured values of absorptivity by Trapp et al. \cite{Trapp2017In} much more closely. Further work is needed on deriving absorptivity scaling laws for high reflectivity aluminium and copper alloys, as the first attempt at in situ absorptivity measurements during LPBF of copper also suggest a significant difference when compared to low reflectivity materials \cite{gargalis2021determining}. 

\begin{figure}[htbp]
    \centering
    \captionsetup{justification=centering}
    \includegraphics[width=15cm,keepaspectratio]{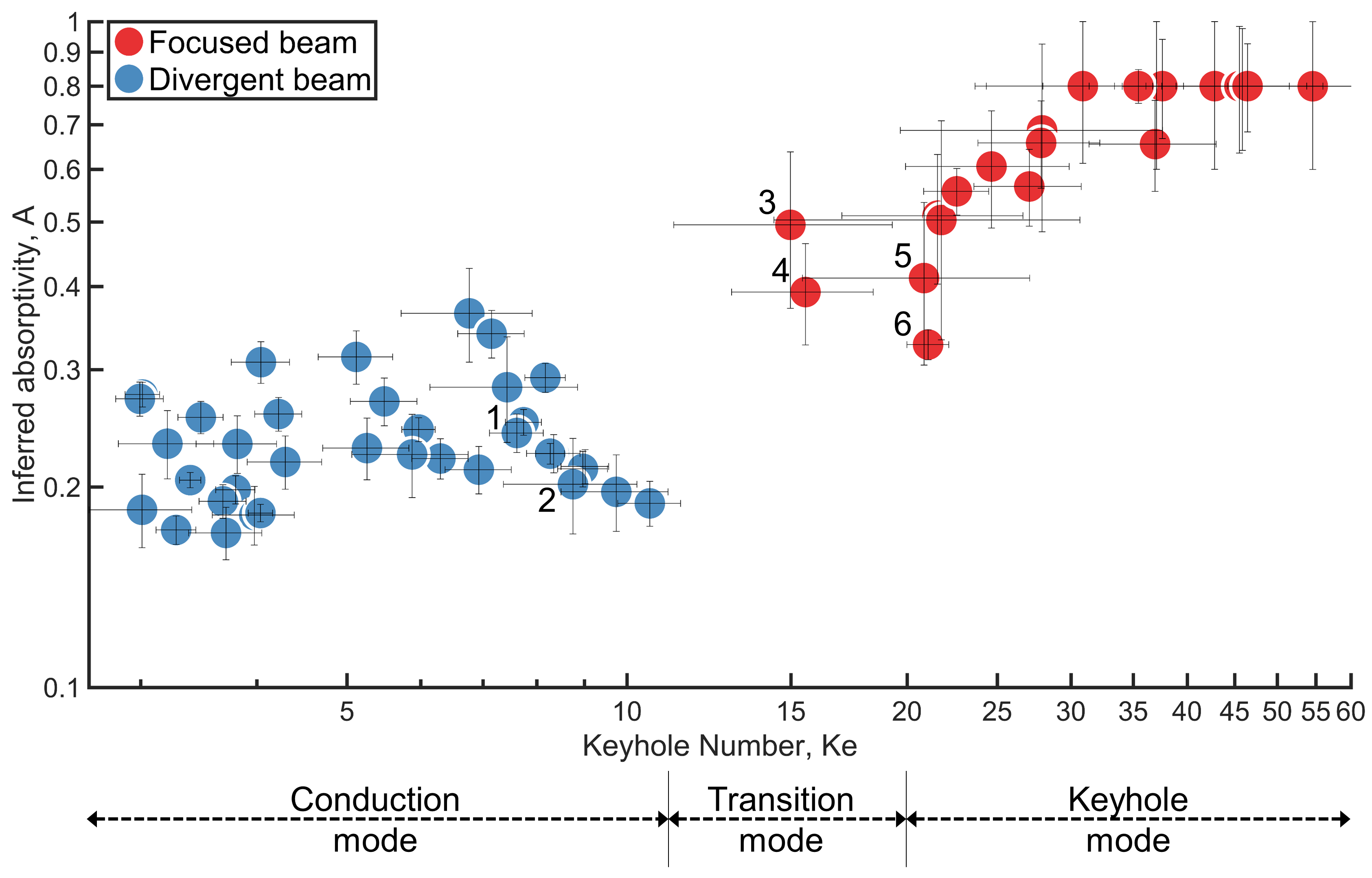}
    \caption{Keyhole number versus inferred absorptivity plot for the 57 AlSi10Mg process parameters used in this study grouped by the use of focused and divergent beams. Samples codes 1-6 used for defect space evaluation in sections~\ref{defect_space_micrographs} and \ref{defect_space_XCT} are labelled in the figure.}
    \label{fig:keyhole_number_ALSi10Mg}
\end{figure}

To offset the limitations of using the normalized processing diagram resulting in uncertainty in identifying melting mode thresholds, and thus to better identify the conduction, transition, and keyhole melting modes, a keyhole number ($Ke$) developed by Gan et al. \cite{gan2021universal} is plotted versus the inferred absorptivity values in Figure~\ref{fig:keyhole_number_ALSi10Mg}. As depicted in Figure~\ref{fig:keyhole_number_ALSi10Mg}, the onset of surface vaporization brings about a more pronounced change in laser absorption and thereby in melt pool behaviour for high reflectivity materials such as aluminium alloys. This is because the onset of surface vaporization adds to additional absorptivity (A) of the laser beam in the material that is equal to $1-R^N$, where $R$ is the reflectivity a material, and $N$ is the number of reflections occurring in the vaporized cavity of the melt pool \cite{Graf2015Analytical}. Materials such as aluminium alloys with higher reflectivity values compared to titanium, ferrous, and nickel alloys would thereby be expected to have differences in melt pool behaviour (melt pool dynamics and thereby solidified melt pool geometry) after the onset of surface of vaporization is crossed. This points towards the differences in absorptivity values that were obtained for the conduction mode points using divergent beams and transition and keyhole mode points using focused beams.

Table~\ref{tab:AlSi10Mg_Process_Outcomes} shows that conduction mode melt pools (based on Figure~\ref{fig:process_diagram_A_F_ALSi10Mg}) obtained using divergent beams have aspect ratios (melt pool depth/width) of 0.33 ± 0.03, whereas the transition and keyhole mode melt pools obtained using focused beams have aspect ratios of 0.47 ± 0.04. By these melt pool aspect ratios, based on existing literature \cite{patel2020melting, stopyra2020laser, King2014Observation, jadhav2021laser, tenbrock2020influence}, it would mean that all the 57 process parameter combination would be expected to lie in the conduction melting mode; however, the optical micrographs of the numerous samples along the build direction (Z-axis) in Figures~\ref{fig:A_F_ALSi10Mg_Microstructure} and \ref{fig:spatter} shows otherwise. Optical micrographs from most of the process parameter combination that use focused beams shown in Figures~\ref{fig:A_F_ALSi10Mg_Microstructure} and \ref{fig:spatter} reveal numerous rounded defects representative of keyhole instabilities implying that these sets of processing parameters likely lie in the keyhole melting mode. Hence, a melt pool aspect ratio of 0.4 is more representative of the threshold between the conduction and keyhole melting modes during LPBF of AlSi10Mg; this finding further refines on the 0.5 aspect ratio threshold in literature \cite{patel2020melting, stopyra2020laser, King2014Observation, chen2021elucidating}. Keyhole numbers of 0-12 in Figure~\ref{fig:keyhole_number_ALSi10Mg} are predicted to lie in the conduction melting mode (melt pool aspect ratios <0.4). Samples 3 and 4 have keyhole numbers of 12-20 that are predicted to lie in the transition melting mode based on the analysis provided in section~\ref{defect_space_micrographs} and \ref{defect_space_XCT}; this range of 12-20 also matches the regime of "stable keyholes" using ultrahigh-speed synchrotron X-ray imaging during LPBF of Ti-6Al-4V, Al6061, and SS316L in work by Gan et al. \cite{gan2021universal}. Keyhole numbers > 20 are predicted to lie the keyhole melting mode during LPBF of AlSi10Mg; keyhole numbers > 20 also correspond to the "chaotic keyholes" regime in the \hl{work} by Gan et al. \cite{gan2021universal}. The absorptivities value obtained for conduction mode melt pools using divergent beams is 0.24 ± 0.05, whereas absorptivity values of 0.65 ± 0.16  are obtained using a focused beam for the transition and keyhole mode melt pools respectively. 

\subsection{Porous defect outcomes across melting modes based on micrographs} \label{defect_space_micrographs}

\begin{figure}[htbp]
    \centering
    \captionsetup{justification=centering}
    \includegraphics[width=15cm,keepaspectratio]{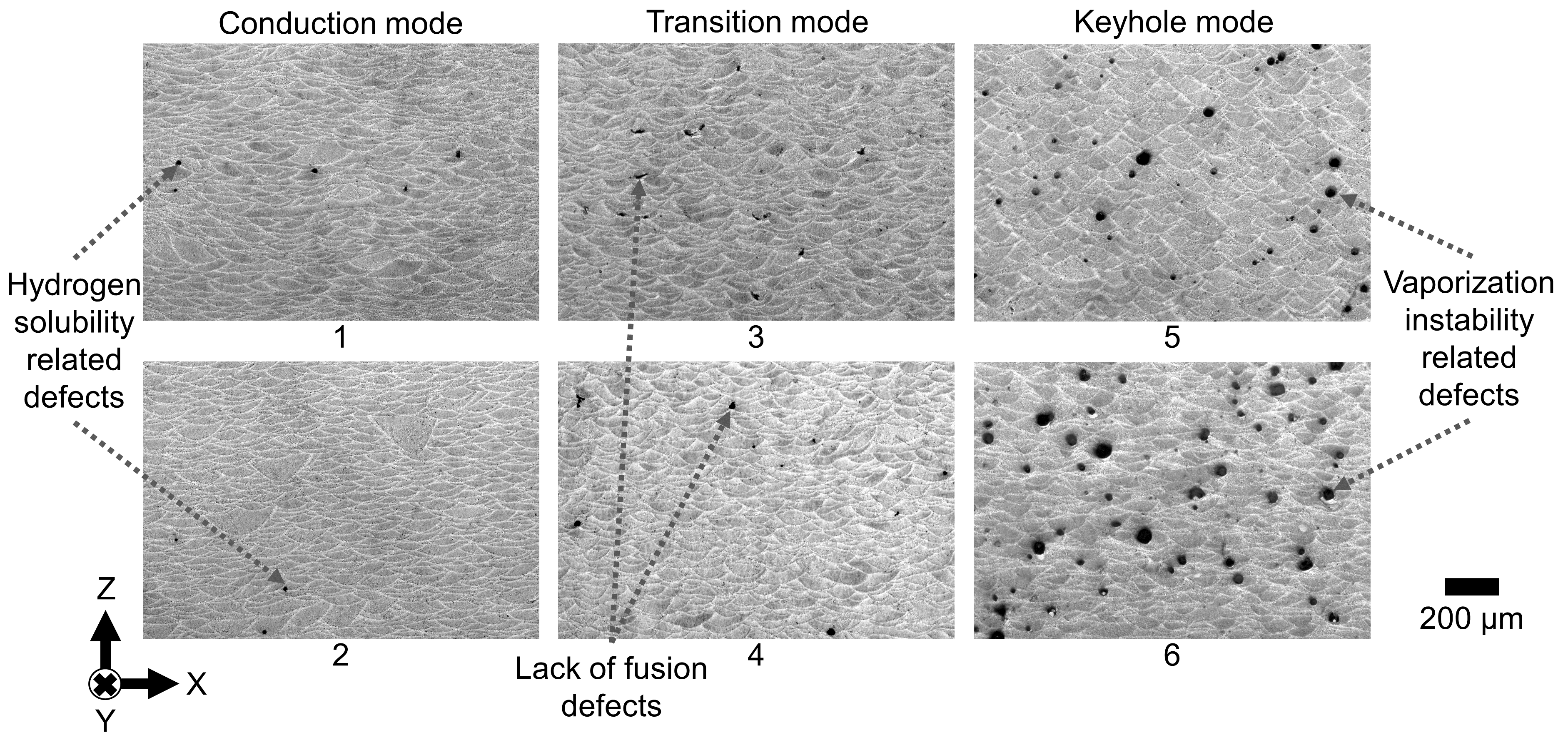}
    \caption{Optical micrographs of the the conduction mode samples (1 and 2), transition mode samples (3 and 4), and keyhole mode samples (5 and 6) along the build direction (Z-axis). Red arrows for sample 4 point to rounded porous defects with diameters of $\sim$25 \textmu m.}
    \label{fig:A_F_ALSi10Mg_Microstructure}
\end{figure}

The effect of the three melting modes (conduction, transition, and keyhole) on melt pool morphologies and porosity was studied using optical micrographs of two conduction mode samples (1 and 2), two transition mode samples (3 and 4), and two keyhole mode samples (5 and 6) along the build direction (Z-axis). Representative illustrations of results are provided in Figure~\ref{fig:A_F_ALSi10Mg_Microstructure}. The transition mode samples (3 and 4) seem to have a similar porosity level to the conduction mode samples (1 and 2) in the microstructural images shown in Figure~\ref{fig:A_F_ALSi10Mg_Microstructure}, but there are qualitative differences between the melt pool morphologies. The qualitative differences between the melt pool morphologies of samples 3 and 4 (transition mode) with respect to samples 1 and 2 (conduction mode) are apparent by virtue of more variability in the melt pool layer-by-layer organization. The quantitative differences between the melt pool morphologies of samples 1 and 2 are better represented in Table~\ref{tab:AlSi10Mg_Process_Outcomes} by the lower standard deviations of their melt pool depths when compared to samples 3, 4, and 5. Additionally, samples 3 and 4 seem to have melt pool depths comparable to sample 1 and 2 even when the laser power settings used for them were close to half, which would imply much lower heat inputs. This is due to the onset of vaporization in these samples due to the focused beam, as predicted in the normalized processing diagram in Figure~\ref{fig:process_diagram_A_F_ALSi10Mg}. 

In terms of porous defects, samples 1 and 2 (conduction mode) have a small population of defects as shown in Figure~\ref{fig:A_F_ALSi10Mg_Microstructure}, with the smallest defects being rounded and commonly attributed to hydrogen-induced defects observed in conduction LPBF of AlSi10Mg; this is also observed by Weingarten et al. \cite{Weingarten2015Formation}. The presence of moisture on the powder surfaces, is one of the main causes attributed to the reduction of hydrogen solubility in aluminium alloys during the resolidification of liquid aluminium \cite{kaplan2017metallurgy}. For samples 5 and 6 (keyhole mode), additional larger defects were observed as seen in Figure~\ref{fig:A_F_ALSi10Mg_Microstructure}, particularly at the bottom of melt pool, close to the melt pool boundaries. The source of these defects is expected to be the excessive vaporization of metal expected in keyhole mode melt pools \cite{patel2020melting}. In conduction mode LPBF of aluminium, where significant vaporization is not expected, the measured absorptivity values for LPBF were $\sim$0.15 for a beam spot diameter of 60 ± 5 \textmu m  \cite{Trapp2017In, Matthews2018Direct}. However, the high reflectivity in such materials would be expected to aid the overall absorptance significantly once vaporization initiates due to increased number of reflections of the laser beam inside the vaporized region, as observed for aluminium discs in transition mode \cite{Trapp2017In}. High-speed and high-resolution X-ray imaging of two aluminium alloys (AlSi10Mg and Al6061) during LPBF has shown that fluctuations in their vaporized areas of melt pools lead to instabilities and thereby to the formation of porous defects, even with a shallow depth of the vaporization regions in keyhole mode melting due to an increased number of laser beam reflections in the melt pool \cite{Hojjatzadeh2020Direct, gan2021universal}. \hl{Defect generating instances within such melt pools with vaporized regions have also been proven to correspond with a reduced laser absorption at the instance using real time laser absorption measurements alongside melt pool geometry visualization using high-speed synchrotron x-ray imaging by Simonds et al.} \cite{simonds2021causal}. A few of the excessive vaporization-related defects are also observed in sample 4 (transition mode), as pointed by the white arrows in Figure~\ref{fig:A_F_ALSi10Mg_Microstructure}. Overall, the micrographs support most of the simulation predictions in the normalized process diagram (Figure~\ref{fig:process_diagram_A_F_ALSi10Mg}) and keyhole number plot (Figure ~\ref{fig:keyhole_number_ALSi10Mg}) of samples 1 and 2 being in the conduction mode and samples 5 and 6 being in the keyhole mode. Section~\ref{defect_space_XCT} describes another simulation approach and additional results from X-ray computed tomography to confirm that samples 3 and 4 lie in the transition melting mode. Dimensionless processing diagrams used in the present work are thereby useful tools for reducing the need for iterative design of experiments as they can help with process parameter planning as well as interpreting the physical origin of porous defects during LPBF.  

\subsection{Porous defect outcomes across melting modes based on XCT} \label{defect_space_XCT}
\begin{figure}[htbp]
    \centering
    \captionsetup{justification=centering}
    \includegraphics[width=16cm,keepaspectratio]{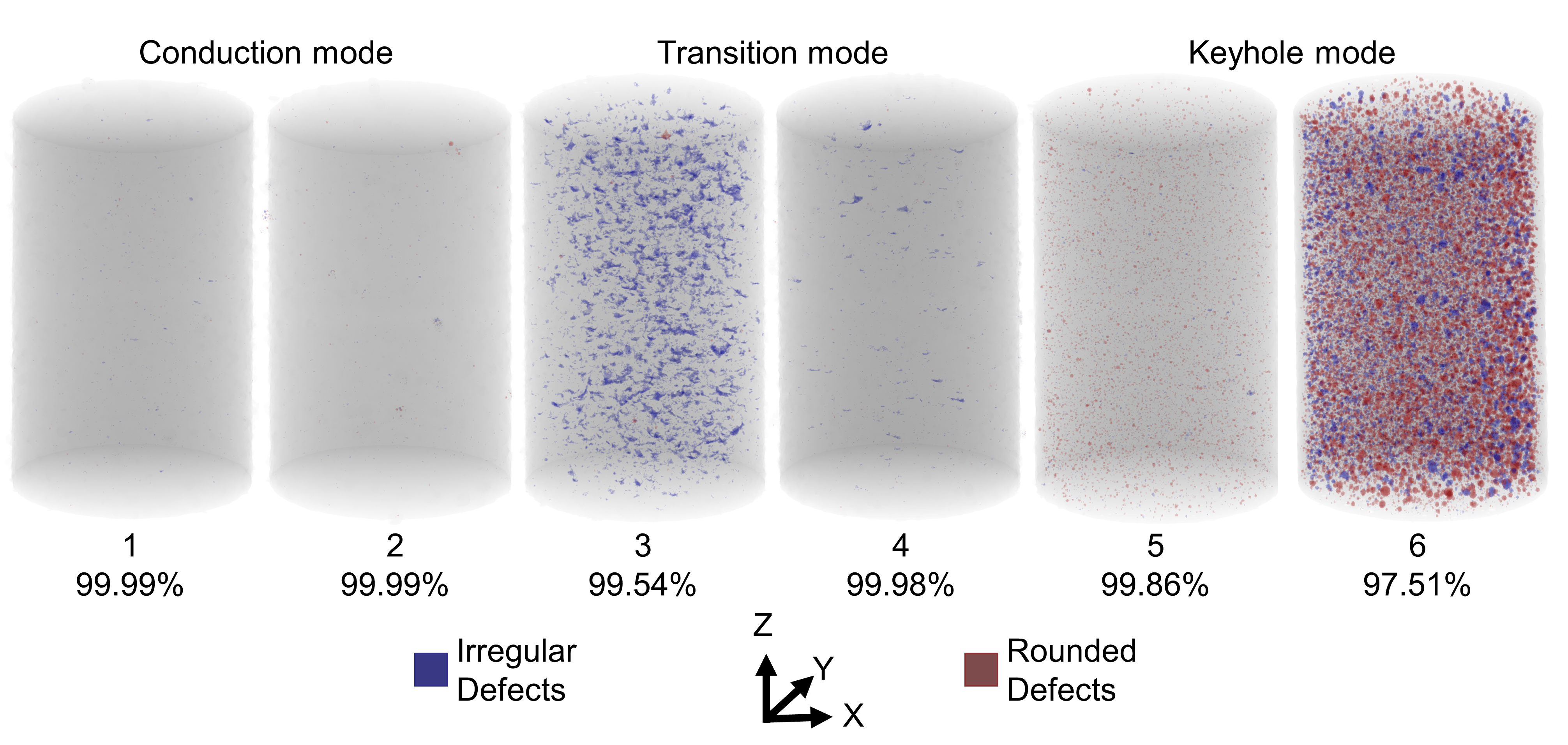}
    \caption{A three-dimensional visualization of the porous defect (above 4 voxels) space along the build direction (Z-axis) from the XCT data of the conduction mode samples (1 and 2), transition mode samples (3 and 4), and keyhole mode samples (5 and 6), along with the density values obtained based on the XCT data. Defects with aspect ratios above 0.7 were considered as rounded defects.}
    \label{fig:XCT_3D_A_F_ALSi10Mg}
\end{figure}

\begin{figure}[htbp]
    \centering
    \captionsetup{justification=centering}
    \includegraphics[width=16cm,keepaspectratio]{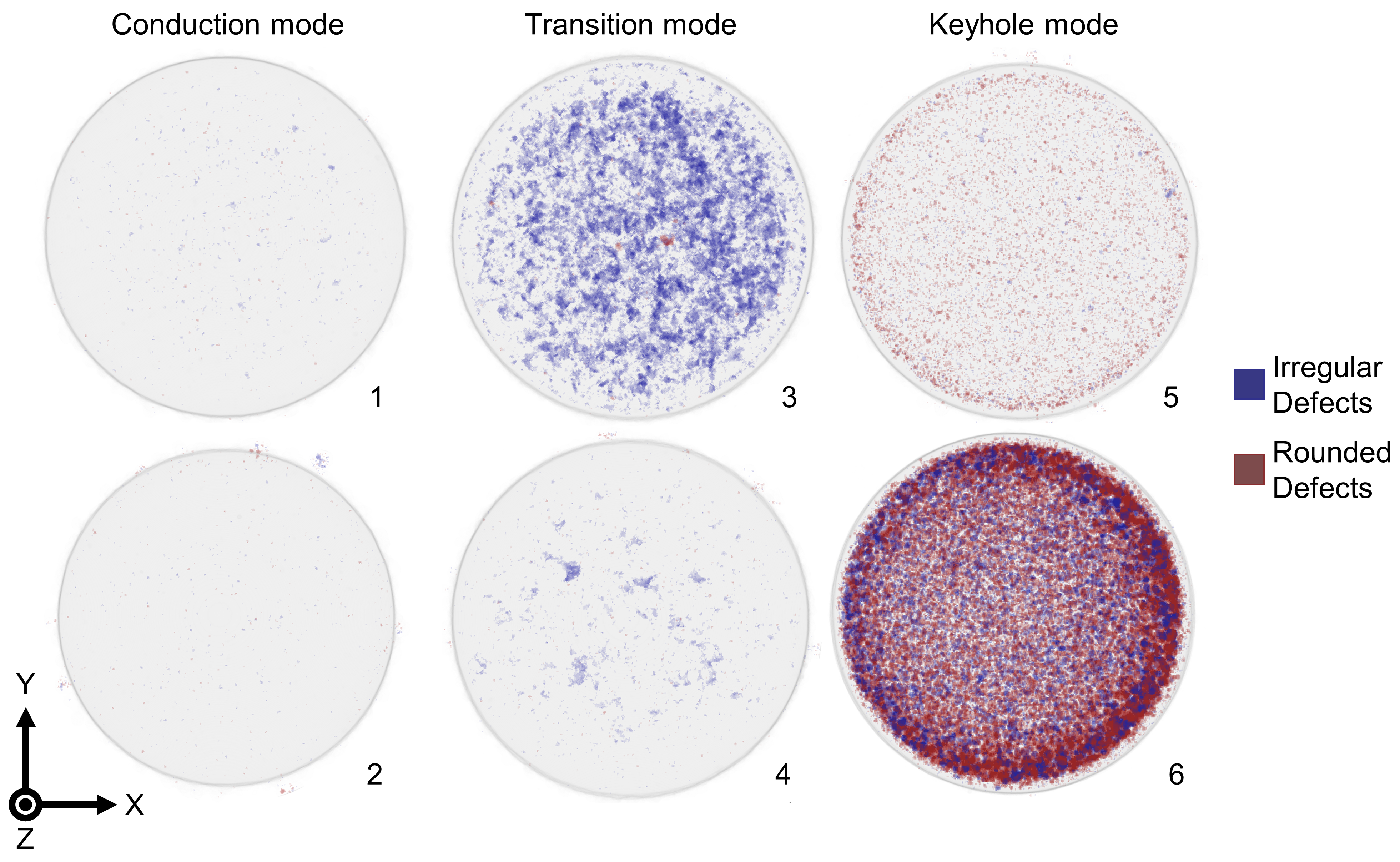}
    \caption{An orthographic projection of the porous defect (above 4 voxels) space along the build plate (XY) plane from the XCT data of the the conduction mode samples (1 and 2), transition mode samples (3 and 4), and keyhole mode samples (5 and 6).Defects with aspect ratios above 0.7 were considered as rounded defects.}
    \label{fig:XCT_Ortho_A_F_ALSi10Mg}
\end{figure}

To further understand the effects of conduction, transition, and keyhole melting modes on defect formation during LPBF of AlSi10Mg, a visualization of the three-dimensional porous defect space (obtained by XCT) for all six samples are shown in Figure~\ref{fig:XCT_3D_A_F_ALSi10Mg}. Segmented defects with sizes below 5 interconnected voxels (voxel edge dimension is 6 \textmu m) have been truncated out from the defect visualization and defect aspect ratio assessments since it is not possible to accurately separate features below this size due to uncertainty in pore segmentation thresholding and instrument noise. The porous defect aspect ratio parameter is the ratio between the minimum and the maximum Feret diameter, where the minimum Feret diameter is the shortest length of a given feature, while the maximum Feret diameter is the longest span of a given feature, as described in \cite{Salarian2020Pore, Asgari2018On}. Defects with aspect ratios above 0.7 were considered as rounded defects in Figures~\ref{fig:XCT_3D_A_F_ALSi10Mg} and \ref{fig:XCT_Ortho_A_F_ALSi10Mg}. For calculating the density values shown in Figure~\ref{fig:XCT_3D_A_F_ALSi10Mg}, all the defects (defects with a voxel size of 1 or more) was considered. The density values are approximations of the true density and a relative assessment of part quality due to the voxel size detection limit. To visualize the locations of the defects, an orthographic projection along the build plate (XY) plane of all the porous defect space for each sample is shown in Figure~\ref{fig:XCT_Ortho_A_F_ALSi10Mg}.

In Figures~\ref{fig:XCT_3D_A_F_ALSi10Mg} and \ref{fig:XCT_Ortho_A_F_ALSi10Mg}, it can be observed that sample 1 (conduction mode) has few irregular defects; these are lack of fusion defects, which are attributed to the slightly lower melt pool depths (when compared to sample 2, conduction mode) as noted in Table 2 \cite{Wang2018processing}. There are numerous causes for lack of fusion defects in conduction mode LPBF such as, but not limited to incomplete melting of powder particles within one-layer, incomplete re-melting of material ejecta from previous layers or from neighbouring scan tracks, or incomplete re-melting of irregular surface topographies from previous layers. Such defects can propagate across subsequent layers, resulting in irregularly shaped lack-of-fusion defects \cite{King2014Observation} that can be observed in Figures~\ref{fig:XCT_3D_A_F_ALSi10Mg} and \ref{fig:XCT_Ortho_A_F_ALSi10Mg}. 

\begin{figure}[htbp]
    \centering
    \captionsetup{justification=centering}
    \includegraphics[width=15cm,keepaspectratio]{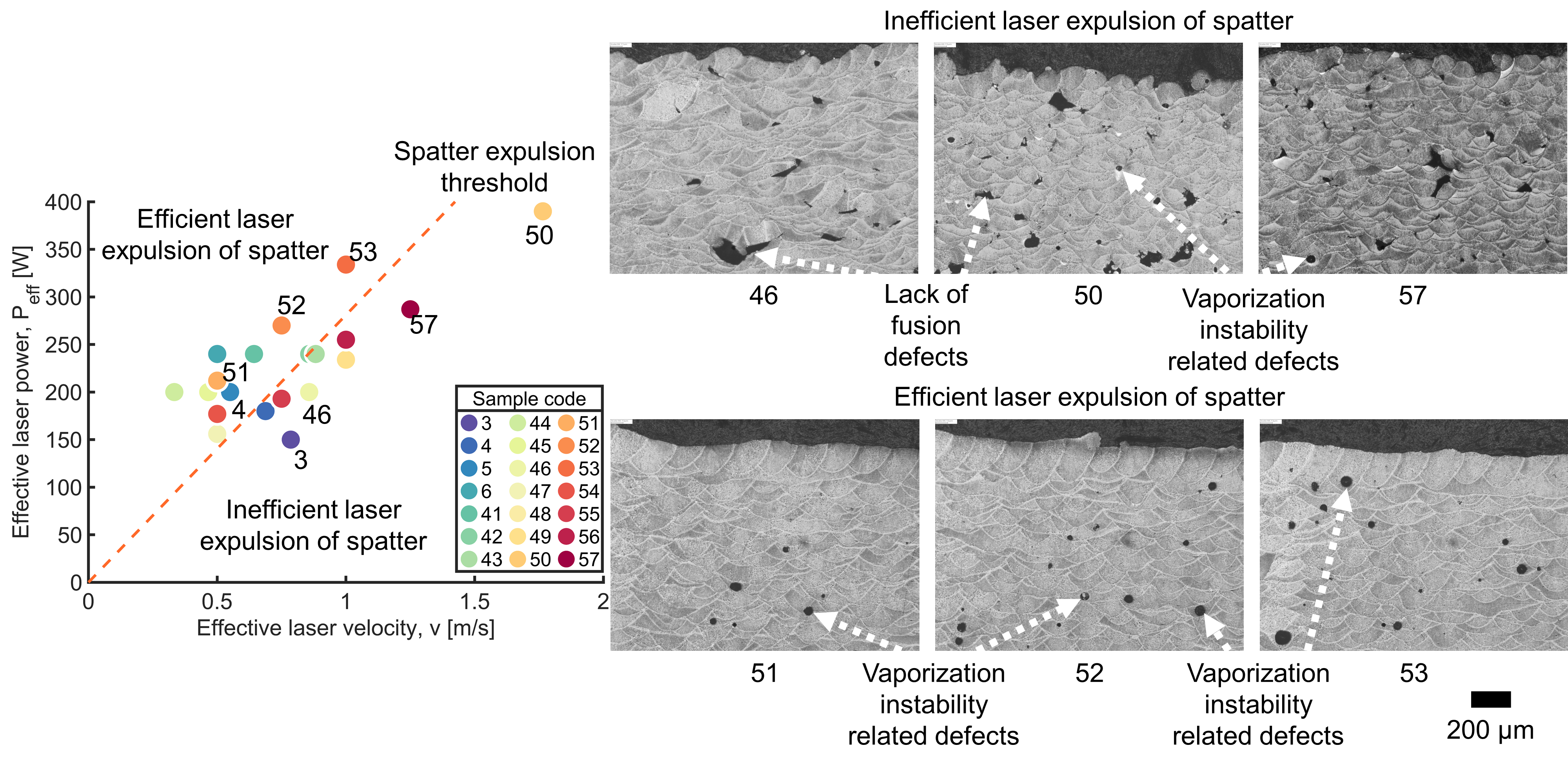}
    \caption{Spatter expulsion window based on effective laser power and velocity for LPBF of AlSi10Mg using a focused beam (beam radius = 35 \textmu m) on the Renishaw AM 400 system (left). Micrographs of 3 sample codes (46, 50, and 57) that lie in the region of inefficient expulsion of spatter, and 3 samples codes (51, 52, and 53) that lie in the region of efficient expulsion of spatter (right).}
    \label{fig:spatter}
\end{figure}

The defect population in Samples 3 and 4 (transition mode) spans both irregularly shaped and rounded (near spherical) defects as seen in Figures~\ref{fig:XCT_3D_A_F_ALSi10Mg} and \ref{fig:XCT_Ortho_A_F_ALSi10Mg}, with a dominance of irregularly shaped defects. While both samples 3 and 4 (transition mode) have similar melt pool dimensions as observed in Table~\ref{tab:AlSi10Mg_Process_Outcomes}, there is a significant difference in the density and frequency of irregularly shaped defects as visible in Figures~\ref{fig:XCT_3D_A_F_ALSi10Mg}. The main cause for the lower density of sample 3 when compared to sample 4 could be associated to the inefficient laser expulsion of metal spatter as predicted by the analytical relationship derived by Khairallah et al. \cite{khairallah2020controlling} in Equation~\ref{P_spatter} and visualized for the 21 focused beam processing parameters from Table~\ref{tab:AlSi10Mg_Process_Para} at a beam diameter of 70 \textmu m for the Renishaw AM 400 system in the left side of Figure~\ref{fig:spatter}.

Khairallah et al. \cite{khairallah2020controlling} used a combination of high-fidelity simulations and high-speed X-ray imaging of LPBF to derive a criteria for stabilizing melt pool dynamics and minimizing defects. They derived an analytical relationship to help identify combinations of laser power and velocity that can help in preventing large metal spatter from blocking the center of a Gaussian laser beam.  In Equation~\ref{P_spatter}, $H$ is the addition of the latent heat of fusion and vaporization, given by total latent heat in Table~\ref{tab:AlSi10Mg_properties} and $A_m$ the laser absorptivity of the melted surface of a given material which is assumed to the 0.15 (lowest conduction mode absorptivity in the present work and experimentally measured by Trapp et al. \cite{Trapp2017In}). The term $r_s$ in Equation~\ref{P_spatter} is the radius of a spatter particle under consideration.

\begin{equation}
    P_{threshold}=\frac{\pi \rho r_s^3 (C_p(T_v-T_0)+H)}{3 A_m r_b}v
   \label{P_spatter}
\end{equation}

Khairallah et al. \cite{khairallah2020controlling} observe that a spatter particle that is as large as the laser beam is capable of blocking the central high intensity region of a Gaussian laser beam, leading to sudden drop in melt pool depth. The rapid cooling caused by the sudden drop in melt pool depth thereby leads to defects as also observed by Martin et al. \cite{Martin2019Dynamics}. The radius of the spatter particle ($r_s$) used to derive the spatter expulsion threshold through Equation~\ref{P_spatter} for Figure~\ref{fig:spatter} is hence assumed to be the same as the beam radius (35 \textmu m) used for the 21 samples codes that used a focused beam in Figure~\ref{fig:spatter}. The exact relationship used for Figure~\ref{fig:spatter} is given by Equation~\ref{P_threshold_contour}.

\begin{equation}
    P_{threshold}=281*v
   \label{P_threshold_contour}
\end{equation}

Inefficient expulsion of spatter during LPBF of AlSi10Mg is confirmed to be a cause for lack of fusion defects in the transition and keyhole mode parameters (that use a focused beam) through representative micrographs on the right side of Figure~\ref{fig:spatter}. Sample codes 46, 50, and 57 (keyhole mode) are similar to sample code 3 (transition mode) in that they lie in the region wherein an inefficient laser expulsion of spatter is expected. This leads to a dominance of irregularly shaped lack of fusion defects, in line with the findings of Khairallah et al. \cite{khairallah2020controlling}. A few rounded vaporization instability related defects are also observed in samples 46, 50, and 57 which would be expected due to the prediction of the keyhole melting mode for these samples in Figures~\ref{fig:process_diagram_A_F_ALSi10Mg} and \ref{fig:keyhole_number_ALSi10Mg}. Sample 4 (transition mode) which shows very few lack of fusion defects lies very close to the spatter expulsion threshold. Sample codes 51, 52, and 53 (keyhole mode) have no lack of fusion defects in Figure~\ref{fig:spatter}. Samples 51, 52, and 53 are similar to sample 4 such that they all lie in the region wherein an efficient expulsion of spatter is expected as per the left side of Figure~\ref{fig:spatter}. The primary source of defects in samples 51, 52, and 53 in the right side of Figure~\ref{fig:spatter} is observed to vaporization instability driven which is expected due to the prediction of the keyhole melting mode for these samples in Figures~\ref{fig:process_diagram_A_F_ALSi10Mg} and \ref{fig:keyhole_number_ALSi10Mg}. Metal vaporization during LPBF was observed to be the largest driver of spatter issues by Khairallah et al. \cite{khairallah2020controlling} and hence the divergent beam sample codes were not plotted in Figure~\ref{fig:spatter} as they would not be affected by vaporization driven spatter related challenges, as predicted by Figure~\ref{fig:process_diagram_A_F_ALSi10Mg}. For practical applications, the spatter expulsion threshold plotted in Figure~\ref{fig:spatter} supplements the normalized processing diagram (Figure~\ref{fig:process_diagram_A_F_ALSi10Mg}) and the keyhole number plot (Figures~\ref{fig:keyhole_number_ALSi10Mg}) by helping predict focused beam LPBF process parameter combinations that would help avoid deleterious lack of fusion defects by virtue of inefficient expulsion of laser spatter. The normalized processing diagram in Figure~\ref{fig:process_diagram_A_F_ALSi10Mg} helps in predicting the extent of vaporization for a given set of process parameters, and the keyhole number plot in Figure~\ref{fig:keyhole_number_ALSi10Mg} helps in identifying the thresholds for the conduction, transition, and keyhole melting modes in LPBF.

There are additional reasons for the defects observed in the transition mode samples 3 and 4. Powder ejecta \cite{Nassar2019Formation} and melt pool splatter caused by the breaking up of elongated molten pool regions near the side and real walls of the vaporized region in transition and keyhole melting modes also contribute towards defect formation in samples 3 and 4 by adding to the roughness of a given layer \cite{Khairallah2016Laser}. Additionally, powder denudation \cite{Matthews2016Denudation, Bidare2018Fluid} also contribute towards the surface roughness of a given layer, thereby deteriorating the wetting behaviour of the following layers causing melt pool instabilities and increasing the possibility of irregularly shaped defects as observed in the defect space visualization for samples 3, 4, 5, and 6 \cite{Zhou20153D-imaging}. Although the average melt pool depth obtained for samples 3 and 4 (transition mode) is similar to sample 1 (conduction mode), the higher standard deviation in melt pool depths observed for samples 3 and 4 (Table~\ref{tab:AlSi10Mg_Process_Outcomes}) can lead to random regions where under-melting may occur if the melt pool is too shallow leading to irregularly shaped lack of fusion defects, or random regions where the process transitions into the keyhole melting mode leading to keyhole defects (sample 4 in Figure~\ref{fig:A_F_ALSi10Mg_Microstructure}). The average melt pool widths of samples 3 and 4 are also lower when compared to samples 1 and 2, leading the possibility of lack of fusion defects caused by insufficient stitching of melt pools between hatches (hatch distance 100 \textmu m) in a given layer. Additionally, as per Figure~\ref{fig:process_diagram_A_F_ALSi10Mg}, it is predicted that surface vaporization has taken place in samples 3, 4, 5, and 6 leading to the possibility of defects related to the melt pool instabilities during transition melting mode in LPBF caused by the interplay between the drag force induced by the melt flow, the thermo-capillary force cause by the surface temperature gradients, and the recoil pressure introduced by the onset of material vaporization \cite{Khairallah2016Laser, Hojjatzadeh2019Pore, Martin2019Dynamics}. The defects obtained due to melt pool instabilities are known to have both rounded and irregularly-shaped morphologies \cite{patel2020melting, Patel2019Melting}. In a comparative study between the laser welding of an aluminium alloy and a ferrous alloy, the higher frequency of vaporized region collapse for aluminium alloys has been associated to the lower surface tension and viscosity of molten aluminium along with the presence of volatile magnesium which vaporizes a temperature much lower than that of aluminium \cite{Huang2018Numerical}. These observations are in line with the hypothesis proposed by Tenbrock et al. \cite{tenbrock2020influence}, wherein the importance of keyholes as a quasi-black body might be more pronounced for materials with higher reflectivity (e.g. aluminium and copper alloys) when compared to titanium, ferrous, and nickel alloys. 

\begin{figure}[htbp]
    \centering
    \captionsetup{justification=centering}
    \includegraphics[width=15cm,keepaspectratio]{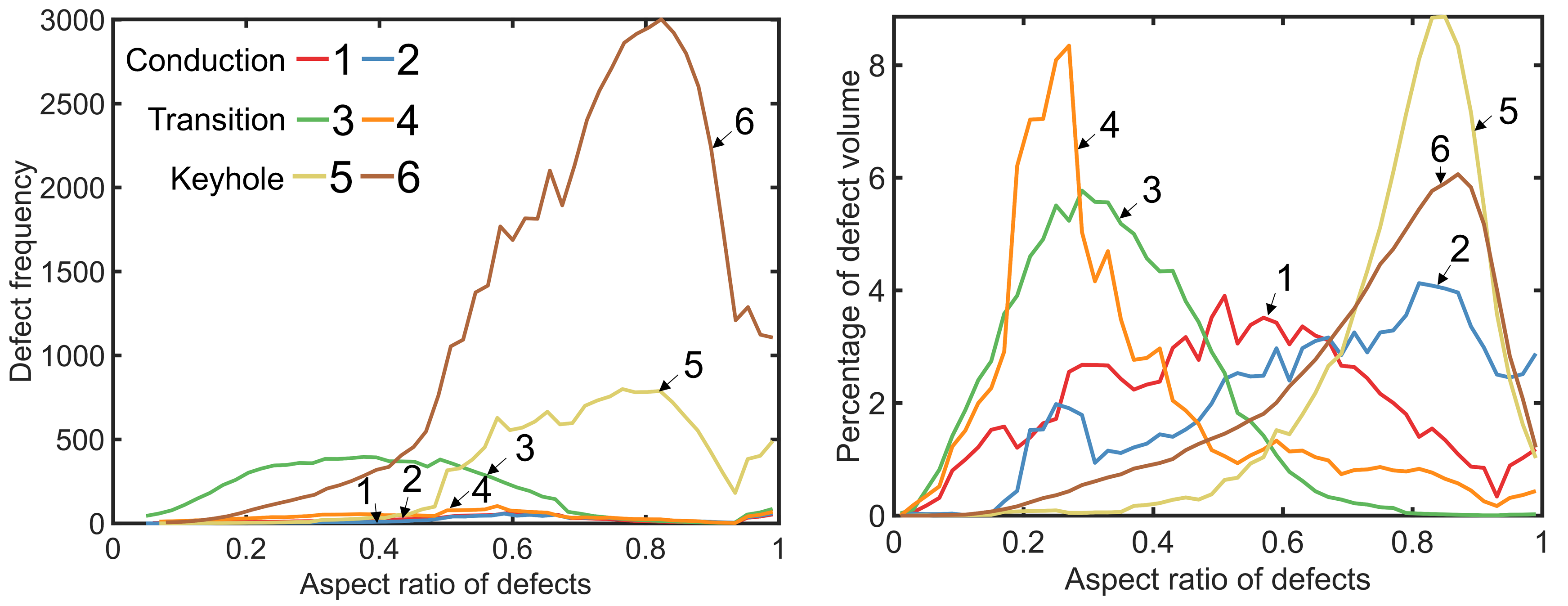}
    \caption{Aspect ratio of defects versus frequency of defects (a), and aspect ratio of defects versus percentage of defect volume (b) from the XCT data for sample codes 1-6 used for defect space evaluation in sections~\ref{defect_space_micrographs} and \ref{defect_space_XCT}. Samples 1 and 2 are predicted to lie in the conduction melting, 3 and 4 in the transition melting mode, and 5 and 6 in the keyhole melting mode.}
    \label{fig:XCT_Aspect_Ratio}
\end{figure}

\begin{figure}[htbp]
    \centering
    \captionsetup{justification=centering}
    \includegraphics[width=8cm,keepaspectratio]{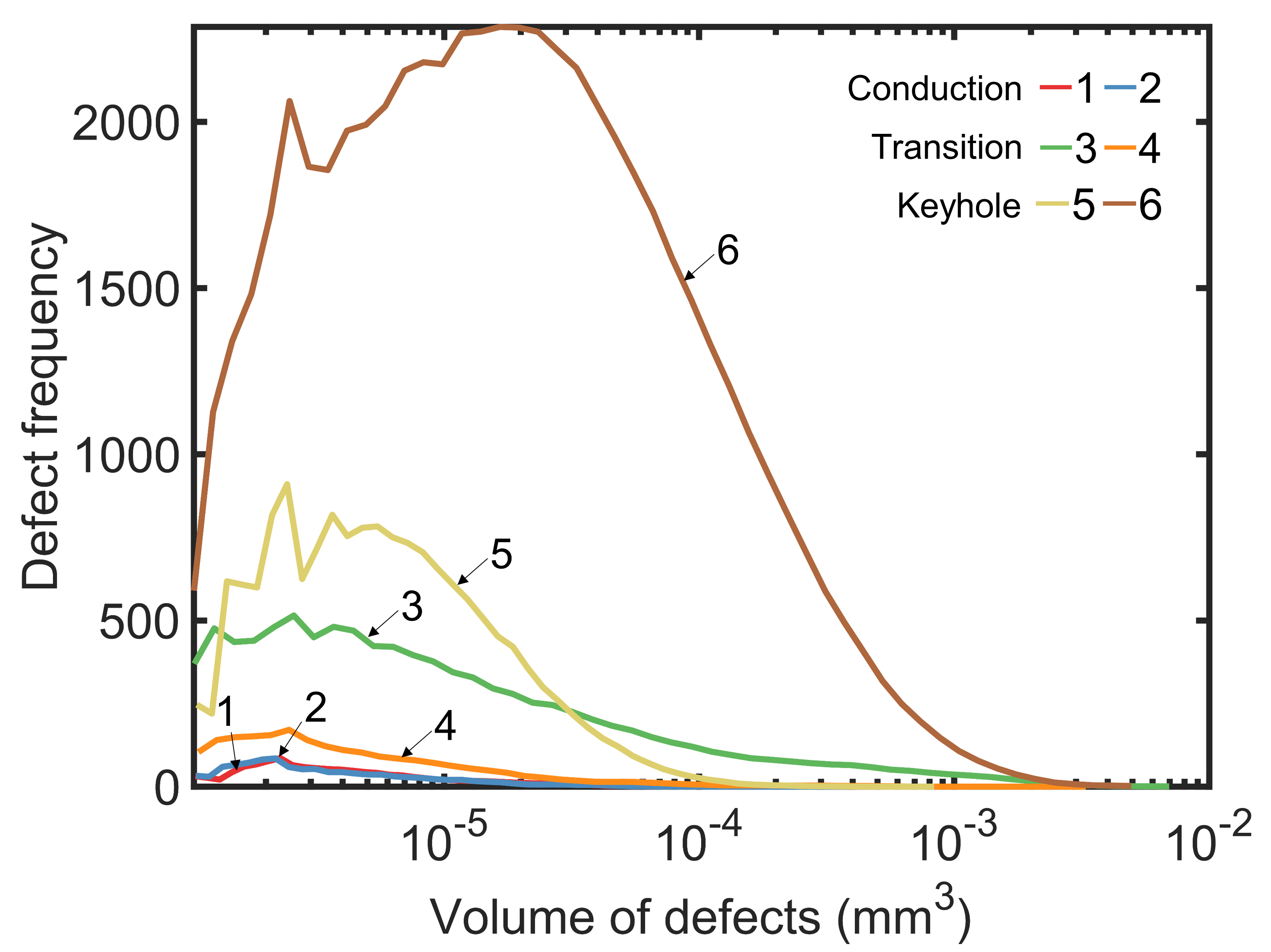}
    \caption{Volume of defects versus frequency of defects from the XCT data for sample codes 1-6 used for defect space evaluation in sections~\ref{defect_space_micrographs} and \ref{defect_space_XCT}. Samples 1 and 2 are predicted to lie in the conduction melting, 3 and 4 in the transition melting mode, and 5 and 6 in the keyhole melting mode.}
    \label{fig:XCT_Volume_freq}
\end{figure}

Samples 5 and 6 are predicted to lie in the keyhole melting mode by Figures~\ref{fig:process_diagram_A_F_ALSi10Mg} and \ref{fig:keyhole_number_ALSi10Mg}. In the keyhole melting mode, vaporization related instabilities inside the melt pool would be expected to play a dominant role in the formation of rounded porous defects, as observed in Figures~\ref{fig:XCT_3D_A_F_ALSi10Mg} and \ref{fig:XCT_Ortho_A_F_ALSi10Mg}. To confirm this prediction, Figure~\ref{fig:XCT_Aspect_Ratio} shows plots of aspect ratios of the defects versus frequency of defects and percentage of defect volume. The term frequency herein means the number of defects within a given sample identified by XCT. The aspect ratio data of the defects in Figure~\ref{fig:XCT_Aspect_Ratio} and the volume data in Figure~\ref{fig:XCT_Volume_freq} is stored in 50 equally sized bins using MATLAB's built-in ‘histcounts’ function. A moving average of the frequency data with 5 nearest neighbours is then calculated using MATLAB's built-in ‘movmean’ which alongside the mid-point of each bin is used to interpolate the curves in Figure~\ref{fig:Asp_freq} and ~\ref{fig:XCT_Volume_freq} by using MATLAB's built-in ‘plot’ function. Figure~\ref{fig:Asp_def_vol} was similarly obtained by creating a moving average of the volume data with respect to aspect ratio of the defects obtained from the XCT data.

The defect aspect ratio versus frequency plot in Figure~\ref{fig:Asp_freq} shows some indication to the preference of rounded defects in samples 5 and 6 (keyhole mode), but the aspect ratio versus percentage of defect volume plot provides a better understanding of such behaviour. Since the curves for samples 5 and 6 (keyhole mode) lean towards a higher aspect ratio in the plot against percentage of defect volume in Figure~\ref{fig:Asp_def_vol}, it implies that most of the defects in samples 5 and 6 have a rounded morphology especially when compared to the other four samples. The rare occurrence of irregularly shaped defects can be associated to the higher average melt pool depths reported in Table~\ref{tab:AlSi10Mg_Process_Outcomes}, which are above two times the powder layer thickness used (30 \textmu m). Typically, a melt pool depth of about 2 times (or more) the layer thickness is targeted in LPBF to avoid the possibility of lack of fusion defects \cite{Wang2018processing}. 

Figure~\ref{fig:XCT_Ortho_A_F_ALSi10Mg} shows a higher concentration of rounded porous defects near the side walls of the cylinders for samples 5 and 6 (keyhole mode) that can be related to the rapid formation and collapse of deep vaporized regions due to the laser beam velocity at the turn points which occurs at the edges of a given layer in LPBF, thereby trapping the atmospheric gas in the solidified part \cite{Martin2019Dynamics}. A plot of defect volume versus frequency in Figure~\ref{fig:XCT_Volume_freq} reveals that most of the defects in samples 5 and 6 still belong to the lower volume regions of below 0.0001 \si{mm^3}. However, defects closer to the side wall of the cylinders would still be expected to impact its fatigue life, since the larger subsurface defects are the biggest factor impacting a parts fatigue life. The largest defect in sample 6 (keyhole mode) has a volume of 0.0053 \si{mm^3}. If this largest defect is assumed to be spherical, we would get a defect diameter of 0.22 mm which is close to the defects diameter of $\sim$0.2 mm that led to the fatigue crack initiation in AlSi10Mg as shown by Plessis et al. \cite{Plessis2020Effects}. 

Samples 1 and 2 which are predicted to lie below the surface vaporization threshold (conduction melting mode) in Figures~\ref{fig:process_diagram_A_F_ALSi10Mg} and \ref{fig:keyhole_number_ALSi10Mg} seem to have almost no subsurface defects, with the sporadic occurrence of small defects typical to LPBF caused by random process factors or systematic machine biases \cite{patel2020melting}. Since no particular pattern is observed in the defect space for samples 1 and 2 from Figure~\ref{fig:A_F_ALSi10Mg_Microstructure}, Figure~\ref{fig:XCT_3D_A_F_ALSi10Mg}, Figure~\ref{fig:XCT_Ortho_A_F_ALSi10Mg}, Figure~\ref{fig:XCT_Aspect_Ratio}, and Figure~\ref{fig:XCT_Volume_freq}, the process parameter combinations involving the use of a divergent beam for samples 1 and 2 might be best suited for near fully dense LPBF AlSi10Mg components, particularly for systems with a smaller beam spot radius at the focal point.

Overall, when observing the melt pool morphology and porous defect characteristics across the melting modes in conduction (samples 1 and 2), transition (samples 3 and 4) and keyhole (samples 5 and 6), the benefit of deploying divergent beams in materials with a high reflectivity and high thermal conductivity becomes apparent. For these material systems, the use of a divergent beam as an energy source results in a more stable melt pool morphology, lower occurrence of porous defects in the core and sub-surface regions, and overall lower porous defect volumes. Furthermore, for such approaches, the hatch spacing, and power levels can be further optimized to minimize lack-of fusion random defects.  For this class of material systems, while the use of a focused beam increases the challenge of finding stable melting process parameter combinations, the use of processing diagrams and dimensionless keyhole numbers as proposed in this work, alongside models for predicting the efficient expulsion of laser spatter can help identify stable melt pool morphologies, similar to the ones with divergent laser beams. It is important to note that the modelling and experimental approaches in this work would be particularly applicable to aluminium alloys that are not prone to cracking during LPBF such as AlSi10Mg, A357, Al-12Si, AlSi7Mg, Scalmalloy, etc \cite{kusoglu2020research}. Conventional high strength aluminium alloys from the 2000 (e.g. Al2024 \cite{del2022cracking}), 6000 (e.g. Al6061 \cite{uddin2018processing, loh2014selective}), and 7000 (e.g. Al7075 \cite{martin20173d}) series are geared towards manufacturing using forging techniques, and are prone to hot (solidification and liquation) cracking cracking during LPBF; the behaviour of such alloys may not be fully predicted using the approaches highlighted in this present work. These alloys require approaches that can reduce the solidification stress as well as the ratio of time spent by molten pool duration in the vulnerable zone (solid fraction between 0.9 and 0.99) to relaxation zone (solid fraction between 0.4 and 0.9) \cite{mondal2022crack}. This has been accomplished previously in literature by either using compatible nanoparticle driven control of solidification \cite{martin20173d}, preheating the LPBF build plate \cite{uddin2018processing}, and/or using a high powered laser beam ($\sim$1000 W) \cite{loh2014selective}.

\section{Conclusions}
The influence of focused and divergent Gaussian laser beams on the melt pool dynamics, microstructure, and porous defects during the laser powder bed fusion of a high reflectivity aluminium alloy (AlSi10Mg) is investigated in this work. The key findings are summarized below:
\begin{enumerate}
    \item 
    Divergent beams help in avoiding keyhole defects by reducing the effective beam power density as the melt pool formation progresses, thereby leading to parts with densities of over 99.98\%, with close to no porous defects in the subsurface regions, by conduction mode melting. \hl{Most defects observed in the conduction mode had volumes lower than 0.00001 mm^3. In contrast, keyhole mode melting led to relatively larger defects, with the largest defect having a volume of 0.0053 mm^3. This large keyhole mode defect with an equivalent diameter of 0.22 mm is similar in size to the defect that led to fatigue crack initiation in another study by Plessis et al.} \cite{Plessis2020Effects} on LPBF of AlSi10Mg.
    \item
    Stabilizing melt pool and spatter dynamics in the transition melting mode by using an appropriate laser power and velocity combination can help in minimizing defects and obtaining densities close to 99.98\%, similar to conduction mode densities. 
    \item
    A melt pool aspect ratio (ratio of depth to width) of $\sim$0.4 is observed to be the threshold between conduction and keyhole mode melt pools in AlSi10Mg, which differs from the conventionally assumed melt pool aspect ratio of 0.5.
    \item
    The inferred absorptivity values for conduction mode melt pools (divergent beams) is 0.24 ± 0.05, while absorptivities of 0.65 ± 0.16 are obtained for the transition and keyhole mode melt pools (focused beam), pointing to the significant differences in laser absorptivity following the onset of surface vaporization in aluminium alloys, when compared to titanium, nickel, and ferrous alloys, due to its higher reflectivity.
    \item
    A higher standard deviation is observed in the melt pool depths for the transition mode (18\% to 27\%) and keyhole mode (4\% to 35\%) melt pools, when compared to the conduction mode (3\% to 19\%) melt pools. The predicted absence of vaporization in conduction mode melt pools is a likely cause for the relatively stable melting behaviour, when compared to transition and keyhole mode melt pools.
    \item
    Conduction mode melting during LPBF of AlSi10Mg is expected within keyhole numbers (Ke) of 0-12, transition mode melting is expected within keyhole numbers of 12-20, and keyhole mode melting is expected for keyhole numbers greater than 20.
\end{enumerate}
The application of the methods proposed in this work can help to quickly identify stable conduction mode LPBF processing parameters for high-reflectivity aluminium alloys. The presence of close to no defects even near the boundaries of LPBF components helps increase the confidence of the process for load bearing and mission critical applications in particular.

\section*{Acknowledgements}
Sagar Patel and Mihaela Vlasea appreciate the funding support received from Federal Economic Development Agency for Southern Ontario (FedDev Ontario grant number 081885). In addition, Sagar Patel and Mihaela Vlasea would like to acknowledge the help of Justin Memar for developing the CT analysis code; and the help of Lisa Brock, Hamed Asgari, Allan Rogalsky, Jerry Ratthapakdee, Grace Kurosad, and Henry Ma with the deployment and characterization of builds. Haoxiu Chen and Yu Zou acknowledge the financial support from Natural Sciences and Engineering Research Council of Canada (NSERC Discovery Grant number RGPIN-2018-05731). Haoxiu Chen also acknowledges the China Scholarship Council for a graduate fellowship (No.201906020162).

\bibliography{main.bib}

\end{document}